\documentclass[a4paper]{jpconf}

\usepackage{bm}
\usepackage{amssymb,amsmath}

\usepackage{comment}
\usepackage{ulem}
\usepackage{graphicx}

\newcommand{\CP}{\mathbb{C}P}
\newcommand{\integer}{\mathbb{Z}}

\begin{document}
\title{Statistical nature of Skyrme-Faddeev models in $2+1$ dimensions and normalizable fermions}

\author{Yuki~Amari$^1$,~Masaya~Iida, and Nobuyuki Sawado$^2$}

\ead{$^1$amari.yuki.ph@gmail.com, $^2$sawadoph@rs.tus.ac.jp}
\address{Department of Physics, Tokyo University of Science, Noda, Chiba 278-8510, Japan}

\begin{abstract}
The Skyrme-Faddeev model has planar soliton solutions with target space $\mathbb{C}P^N$. 
An Abelian Chern-Simons term (the Hopf term) in the Lagrangian of the model plays a crucial role
for the statistical properties of the solutions. 
Because $\Pi_3(\mathbb{C}P^1)=\mathbb{Z}$, the term becomes an integer for $N=1$. 
On the other hand, for $N>1$, it becomes perturbative because $\Pi_3(\mathbb{C}P^N)$ is trivial. 
The prefactor $\Theta$ of the Hopf term is not quantized, and its value depends on 
the physical system. 
We study the spectral flow of the normalizable fermions coupled with the 
baby-Skyrme model ($\mathbb{C}P^N$ Skyrme-Faddeev model). We discuss whether
the statistical nature of solitons can be explained using their constituents, i.e., the quarks. 
\end{abstract}

\section{Introduction}

The Skyrme-Faddeev model is an example of a field theory that admits finite-energy knotted solitons. 
The classical soliton solutions of the Skyrme-Faddeev model 
can be useful for describing the strong-coupling sector of the Yang-Mills theory.  
It was shown in \cite{Ferreira:2010jb} that in the case of the complex projective target space $\mathbb{C}P^N$,  
a Skyrme-Faddeev-type model has an infinite number of exact soliton solutions in the integrable sector if the 
coupling constants satisfy a special relation. 
The existence of vortex solutions of the model outside the integrable sector was confirmed numerically for an 
appropriate choice of potentials~\cite{Amari:2015sva}. 
We note that the model is essentially equivalent to the so-called baby-Skyrme model~\cite{Piette:1994ug}, 
which is a (2+1)-dimensional analogue of the Skyrme model. 

It is well known that from the quantum standpoint, soliton solutions have a special property (have
gfractionalh spin statistics) if the Hopf term (theta term) is included in the action of the model~\cite{Wilczek:1983cy}. 
Because $\Pi_3 (\mathbb{C}P^1)=\mathbb{Z}$, this term becomes the Hopf invariant and can therefore be represented as a total derivative,
which does not influence the classical equations of motion. On the other hand, because $\Pi_4(\mathbb{C}P^1)$ is trivial,
the coupling constant $\Theta$ (prefactor of the Hopf term) is not quantized. As shown in~\cite{Wilczek:1983cy}, in the model with
the Hopf Lagrangian, solitons with a unit topological charge have the spin $\frac{\Theta}{2\pi}$, which can be fractional.
For a fermionic model coupled with a $\mathbb{C}P^1$-valued field, $\Theta$ can be found, for example, by using perturbation
theory~\cite{Abanov:2000ea,Abanov:2001iz}.

For $N>1$, the algebra $\Pi_3(\mathbb{C}P^N)$ is already trivial, and the Hopf term is then perturbative, i.e., it
is not a homotopy invariant, and this in the general case means that the contribution from the term can
always be fractional even if we choose an integer $n$ in the anyon angle $\Theta=n\pi$. 
It was pointed out in~\cite{Bar:2003ip}
that an analogue of the Wess-Zumino-Witten term appears in a $\mathbb{C}P^N$-valued field and plays a role similar
to the Hopf term for $N=1$~\cite{Jaroszewicz:1985ip}. 
As a result, the soliton can be quantized as an anyon with the statistics angle $\Theta$ and such a Hopf-like term.

Here, we solve the fermionic model coupled with the baby-Skyrme model or the $\mathbb{C}P^N$ Skyrme-Faddeev
model. The basic property of the localizing mode of fermions on a topological soliton is understood in
terms of a basis from the Atiyah-Singer index theorem~\cite{Atiyah:1968mp}. The index for the Dirac operator $D$ 
can be defined as $\dim\ker D - \dim\ker D^\dagger$, which is related to the Casimir energy of the fermions. 
The spectral flow analysis in the chiral-invariant model (the Skyrme model) is a simple realization of the theorem~\cite{Kahana:1984be}.
When the number $B$ of the one-particle spectra passes from the positive to the negative continuum, the
size or strength of the background skyrmions changes. As a result, the Casimir energy has $B$ states, and
this corresponds to solitons. In the Skyrme model, the Wess-Zumino-Witten term is topological and is
the origin of the topological nature of solitons (skyrmions) and fermions. We therefore believe that the
statistical nature of the soliton is related to the localizing fermions. Because the Hopf term of the $\mathbb{C}P^N$
model is not topological, we expect that the consistency between the statistical nature of the soliton and the
nature of the localizing fermions is broken. Analysis of the spectral flow argument yields new information
about this consistency (or inconsistency).
We consider only the case $N = 2$, but the generalization to large $N$ is straightforward.

\section{The baby-Skyrme model}
\subsection{The model and the toplogical charge}

We introduce the $\mathbb{C}P^1$ model, the so-called baby-Skyrme model. 
The full canonical quantization of this model was already studied in~\cite{Acus:2009df} but without
the Hopf term in the action. There are many studies of fermions coupled with the nonlinear sigma
model~\cite{Abanov:2000ea,Abanov:2001iz,Jaroszewicz:1985ip}, 
including analysis of the spectral flow (see, e.g.,~\cite{Liu:2017rps} for a recent study). 
The Lagrangian of this model is written as
\begin{eqnarray}
{\cal L}_{\rm baby}=M^2\partial_\mu\vec{n}\cdot\partial^\mu\vec{n}
+\frac{1}{e^2}(\partial_\mu\vec{n}\times\partial_\nu\vec{n})\cdot(\partial^\mu\vec{n}\times\partial^\nu\vec{n})
-\mu^2V
\label{lagrangian_bs}
\end{eqnarray}
where $\vec{n}=(n_1,n_2,n_3)$ is constrained to the surface of a sphere of unit radius: $\vec{n}\cdot\vec{n}=1$. 
The positive parameter $M^2$ is a coupling constant with the dimension of mass, and the coupling constant $e^{-2}$ 
has the dimension of inverse mass ($e^2$ must be negative for the Hamiltonian to be positive). The potential term
containing no derivative is denoted by $\mu^2V$ , and $\mu^2$ is the coupling strength. The boundary condition
$\vec{n}_\infty = (0,0,1)$ allows a one-point compactification of the space $\mathbb{R}^2\to S^2$. 
Therefore, skyrmions arise for a map $\vec{n}:S^2\to S^2$. 
This map belongs to an equivalence class characterized by the homotopy group
$\pi_2(S^2)=\mathbb{Z}$. The solutions called baby skyrmions are obtained by solving the Euler-Lagrange equations
for (\ref{lagrangian_bs}) by introducing an appropriate ansatz or by simplifying the Hamiltonian with numerical algorithms,
for example, simulated annealing. For our numerical analysis, we use the standard hedgehog ansatz
\begin{eqnarray}
\vec{n}=(\sin f(r)\cos n\varphi,\sin f(r)\sin n\varphi,\cos f(r))
\label{ansatz_bs}
\end{eqnarray}
with  the boundary condition 
\begin{eqnarray}
f(0)=\pi,~~f(\infty)=0,~~n\in \mathbb{Z}\,.
\label{bc_bs}
\end{eqnarray}

The topological invariant is 
\begin{eqnarray}
Q_{\rm top}=-\frac{1}{4\pi}\int \vec{n}\cdot (\partial_1\vec{n}\times \partial_2 \vec{n})d^2x\,.
\label{tcharge0}
\end{eqnarray}
In terms of the ansatz (\ref{ansatz_bs}) with the boundary condition (\ref{bc_bs}), we easily obtain $Q_{\rm top}\equiv n$. 
To discuss quantization, it is more convenient to use the $SU(2)$-valued field $U:=\vec{\tau}\cdot\vec{n}$. 
We can rewrite the Lagrangian as
\begin{eqnarray}
{\cal L}_{\rm baby}=\frac{M^2}{2}{\rm Tr}(\partial_\mu U\partial^\mu U)
-\frac{1}{8e^2}{\rm Tr}([\partial_\mu U,\partial_\nu U]^2)-\mu^2V
\end{eqnarray}
The topological current is defined in terms of $U$ as
\begin{eqnarray}
j^{\mu}(U)=\frac{i}{16\pi}\epsilon^{\mu\nu\lambda}{\rm Tr}(U\partial_\nu U\partial_\lambda U)
\end{eqnarray}
and the topological charge (\ref{tcharge0}) is expressed by
\begin{eqnarray}
Q_{\rm top}=\frac{i}{16\pi}\int d^2xj^{0}(U)=\frac{i}{16\pi}\int d^2x\epsilon_{ij}{\rm Tr}(U\partial_iU\partial_jU)
\label{tcharg1}
\end{eqnarray}

We again mention that the analogue of the Wess-Zumino-Witten term for baby skyrmions was already
discussed in the literature~\cite{Bar:2003ip,Jaroszewicz:1985ip}. It is given by
\begin{eqnarray}
S_{\rm WZW}^{(3)}=\frac{\Theta}{128\pi^2}\int_{M_4}d^4x\epsilon^{\mu\nu\alpha\beta}{\rm Tr}[U\partial_\mu U\partial_\nu U\partial_\alpha U\partial_\beta U]
\label{wzw}
\end{eqnarray} 
or, a more convenient form easily deduced from (\ref{wzw})
\begin{eqnarray}
S_{\rm WZW}^{(3)}=\Theta\int_{\partial M_4}d^3x {\cal L}_{\rm Hopf},~~~~{\cal L}_{\rm Hopf}:=\frac{1}{4\pi^2}\epsilon^{\mu\nu\alpha}a_\mu\partial_\nu a_\alpha
\label{hopf}
\end{eqnarray}
where $a_\mu:=-iZ^\dagger\partial_\mu Z$. 
These complex coordinates and the field $U$ are related by $U:=1-2Z\otimes Z^\dagger$.
Because $\Pi_4(\mathbb{C}P^1)$ is trivial, the prefactor $\Theta$ does not require quantization. 
This coefficient is sometimes called the anyon angle because 
it is related to the fractional angular momentum of the baby skyrmions.

For the subsequent analysis, we introduce the dimensionless coordinates $(t,\rho,\varphi)$ defined as
\begin{eqnarray}
x^0=r_0t,\quad x^1=r_0\rho\cos\varphi,\quad x^2=r_0\rho\sin\varphi\quad 
\end{eqnarray}
where the length scale $r_0$ is defined in terms of coupling constants $M^2$ and $e^2$, i.e.,
$$
r_0^2=\frac{4}{M^2|e^2|}
$$
and the light speed is $c=1$ in the natural units. The linear element $ds^2$ is
$$
ds^2=r_0^2(dt^2-d\rho^2-\rho^2d\varphi^2).
$$

\begin{figure*}[t]
  \begin{center}
    	\includegraphics[width=90mm]{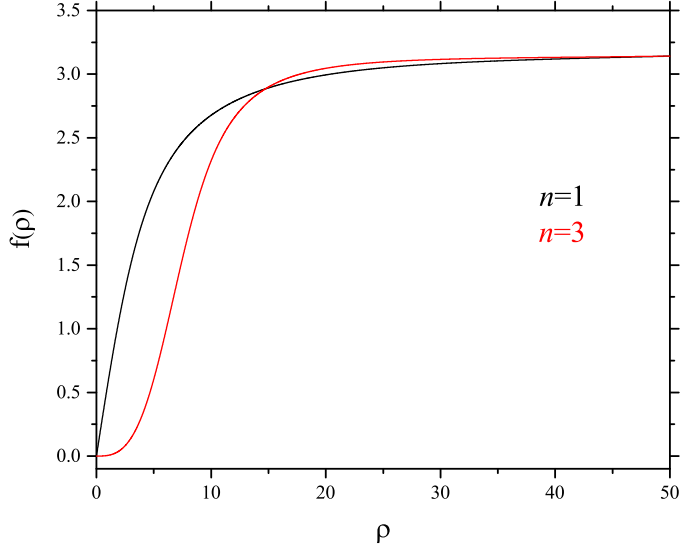}\hspace{-1.0cm}
    \caption{\label{profiles_bs}
The profile $f(\rho)$ of the baby skyrmions $Q_{\rm top} = n, n = 1, 3$: 
the potential is a standard old-baby type $V := 1 -n_3$, 
and the parameter $\tilde{\mu}:=\frac{r_0^2}{M^2}\mu^2$ is chosen equal to 0.005.
}
  \end{center}
	\end{figure*}

\begin{figure*}[t]
  \begin{center}
    	\includegraphics[width=110mm]{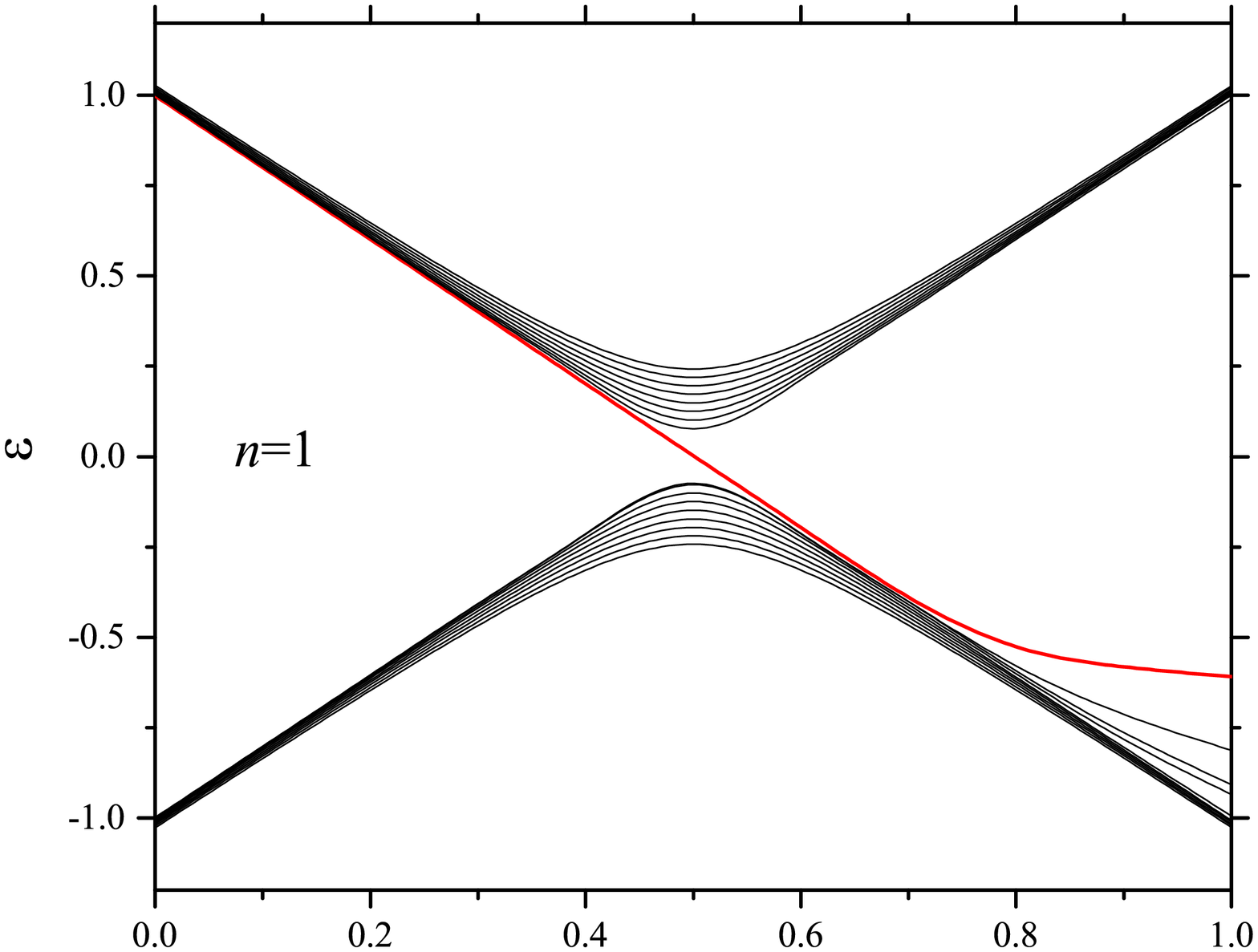}
\vspace{0.5cm}

 	\includegraphics[width=110mm]{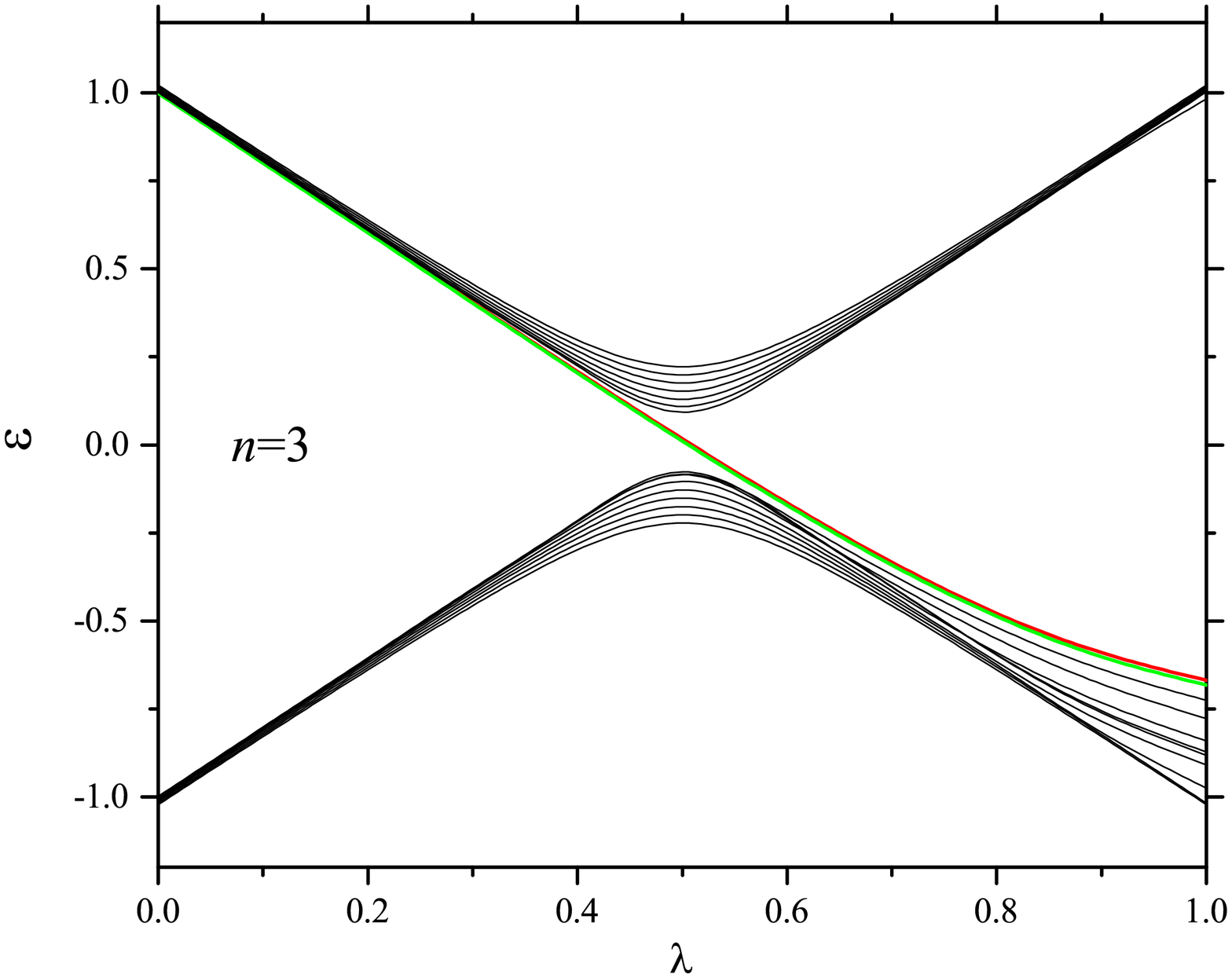}

    \caption{\label{spectralflow_bs} 
The spectral flow corresponding to the solutions shown in Fig. \ref{profiles_bs} for (a) $n = 1$ and (b) $n = 3$:
the background field is the vacuum field $U_\infty$ for $\lambda = 0$ and the soliton field $U$ for $\lambda = 1$. 
We plot the first 18 levels (9 positive and 9 negative energy levels at the vacuum $\lambda=0$). The coupling constant
is chosen as $m = 1.0$, and the potential parameter is $\tilde{\mu}^2=0.005$.
}
  \end{center}
	\end{figure*}

\begin{figure*}[t]
  \begin{center}
    	\includegraphics[width=90mm]{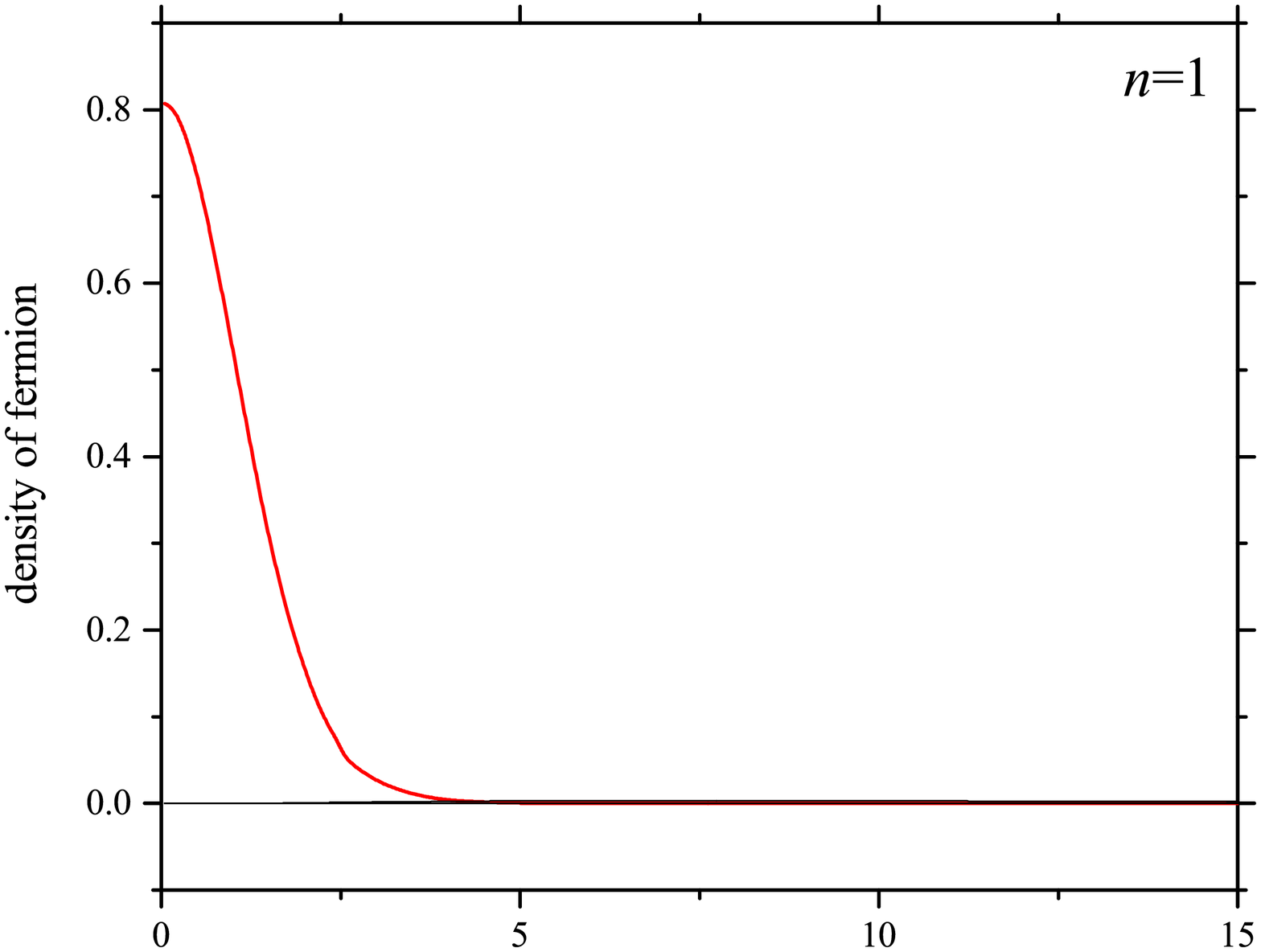}\\
\vspace{-0.0cm}

\hspace{0.8cm}\includegraphics[width=110mm]{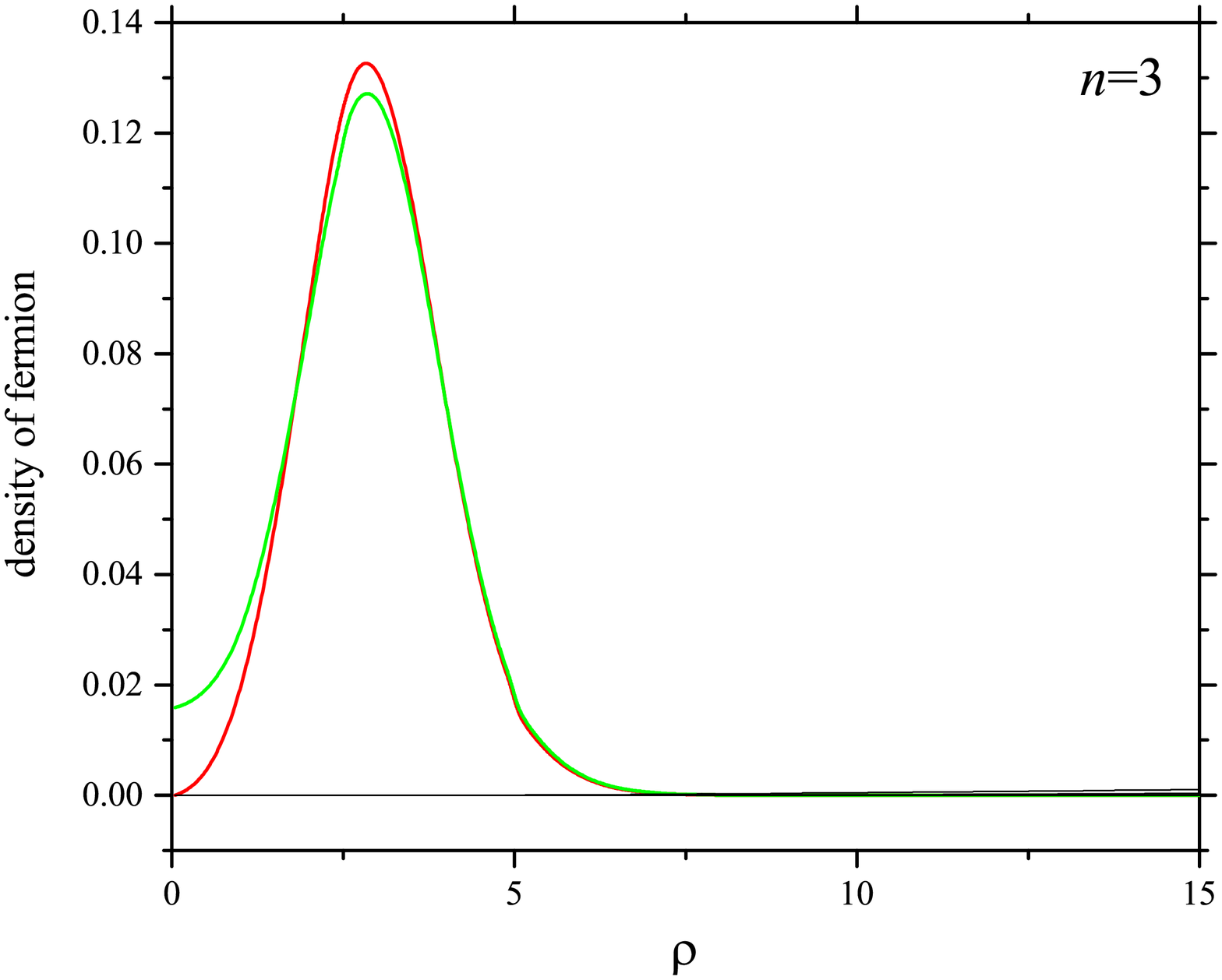}
\caption{\label{fermiondensity_bs}
Fermion density (\ref{density}): the eigenfunction is computed with the soliton background at $\lambda=1$.
The curves correspond to the bold lines in Fig. \ref{spectralflow_bs} (two lines almost merge in Fig. \ref{spectralflow_bs}(b)). 
The potential parameter is $\tilde{\mu}^2=0.1$.
}
  \end{center}
	\end{figure*}

\subsection{The normalizable fermions}

The fermion-vortex system was first studied by Jackiw and Rossi~\cite{Jackiw:1981ee}.
It is well known that a fermionic effective model coupled to the baby skyrmion with a gap $mU$ by integrating
over the fermionic field leads to an effective Lagrangian containing a baby-Skyrme-like model and some
topological terms including the Hopf term~\cite{Abanov:2000ea,Abanov:2001iz}.
We consider a gauged model
\begin{eqnarray}
{\cal L}_{\rm fermi}=\bar{\psi}i\gamma^\alpha(\partial_\alpha-iA_\alpha)\psi-m\bar{\psi}U\psi
\equiv \bar{\psi}i{\cal D}_A\psi
\end{eqnarray}
where m is the coupling constant of baby skyrmions to fermions. The matrices $\gamma_\alpha$ are defined standardly:
$\gamma^1=-i\sigma_1, \gamma^2=-i\sigma_2, \gamma^3=\sigma_3$, where $\sigma_\alpha$ are standard Pauli matrices. 
Under an appropriate rescaling of the Lagrangian, i.e., $\gamma^1=-i\sigma_1, \gamma^2=-i\sigma_2, \gamma^3=\sigma_3$, 
the system becomes dimensionless. The Euclidean partition function is
\begin{eqnarray}
{\cal Z}=e^{\omega(U)}=\int{\cal D}\psi{\cal D}\bar{\psi}\exp(\bar{\psi}i{\cal D}_A\psi)=(\det i{\cal D}_A)^{n_c} .
\label{partitionfunction}
\end{eqnarray}
We separate the effective action $\omega(U)$ into real and imaginary part:
\begin{eqnarray}
&&\omega(U)=n_c{\rm Sp}\ln(i{\cal D}_A)
=\omega(U)_{\rm Re}+i\omega(U)_{\rm Im}\nonumber \\
&&\omega(U)_{\rm Re}=\frac{n_c}{2}{\rm Sp}\ln {\cal D}_A^\dagger {\cal D}_A,
\label{effectiveactionr}\\
&&\omega(U)_{\rm Im}=\frac{n_c}{2i}{\rm Sp}\ln \frac{i{\cal D}_A}{(i{\cal D}_A)^\dagger}
\label{effectiveactioni}
\end{eqnarray}
where ${\rm Sp}$ denotes a full trace containing a functional and also a matrix trace involving the flavor and the
spinor indices and ${\rm Tr}$ denotes the usual matrix trace.
There are many papers devoted to calculating the effective action $\omega(U)=\log {\cal Z}$ in terms of a derivative
expansion. The result contains both the action of the model (in the real part) and the topological terms
(in the imaginary part). After a lengthy calculation (see the appendix), we obtain the effective action in
the Minkowski space-time:
\begin{eqnarray}
&&\omega(U)_{\rm Re}|_{A=0}=n_c\frac{|m|}{16\pi}\int d^3x{\rm Tr}(\partial_\mu U\partial^{\mu} U)
+({\rm higher~derivative~terms}),
\label{effectiveaction_fr}
\\
&&\omega(U)_{\rm Im}=-n_c\int d^3x\Bigl(j^\mu (U) A_\mu+\pi\,{\rm sgn}(m){\cal L}_{\rm Hopf}(U)\Bigr)\,.
\label{effectiveaction_fi}
\end{eqnarray}
where $n_c$ is a degeneracy of the fermions.

If we regard the effective action as the action of a baby-Skyrme-type model, then we can define the
anyon angle $\Theta$ as $\Theta\equiv n_c\pi {\rm sgn}(m)$,, and it is then quantized as usual spin statistics. 
Therefore, similarly to the case of the Skyrme model, the fermion number of the baby skyrmion coincides 
with its topological charge. Indeed, the invariance of $\omega(U)$ under an isosinglet transformation leads to a conserved fermion
current~\cite{Witten:1983tw}. The analysis is perturbative, i.e., the expansion is only justified for small momenta compared
with a physical cutoff.

There is another definition of the fermion number in the skyrmion background related to the distorted
Dirac vacuum~\cite{Diakonov:1987ty,Alkofer:1995mv},
\begin{eqnarray}
N_{\rm cas}:=-\frac{1}{2}\sum_\mu[{\rm sgn}(\epsilon_\mu) -{\rm sgn}(\epsilon_\mu^0)]
\label{ncas}
\end{eqnarray}
where $\epsilon_\mu$ are the eigenvalues of the Hamiltonian
\begin{eqnarray}
i{\cal D}_{A=0}:=\gamma_3(i\partial_3-{\cal H}_{\rm fermi}),~~~~ 
{\cal H}_{\rm fermi}=-i\gamma_3\gamma_k\partial_k+\gamma_3m U.
\end{eqnarray}
and $\epsilon_\mu^0$ is the similar eigenvalue at $U = 1$. The number $N_{\rm cas}$ 
is the Casimir energy, which counts the number of the negative energy levels minus 
those of the vacuum background. Expression (\ref{ncas}) is directly obtained from the 
effective action $\omega (U)$. Therefore, at least within the perturbative regime, both results
should coincide: $N_{\rm cas}\equiv Q_{\rm top}$. 
In what follows, we numerically confirm this coincidence using the spectral flow analysis.

\subsection{The numerical analysis}

We present the typical baby skyrmions with the topological charges $Q_{\rm top}=1,3$ in Fig.\ref{profiles_bs}.
The explicit form of the hamiltonian is 
\begin{eqnarray}
{\cal H}_{\rm fermi}
=
\left(
\begin{array}{cc}
m U & -e^{-i\varphi}\Bigl(\partial_\rho-\frac{i\partial_\varphi}{\rho}\Bigr) \\
e^{i\varphi}\Bigl(\partial_\rho+\frac{i\partial_\varphi}{\rho}\Bigr)& -m U  \\
\end{array}
\right)\,.
\label{hamiltonian_bs}
\end{eqnarray}
It can be shown that the Hamiltonian ${\cal H}_{\rm fermi}$ commutes with the operator of angular momentum, which we
call {\it a grand spin}:
\begin{eqnarray}
{\cal K}=\ell_{3}+\frac{\sigma_3}{2}+\frac{\tau_3}{2}
\label{eqm000}
\end{eqnarray}
where $\ell_3=({\bf r}\times{\bf p})_3=-i\frac{\partial}{\partial\varphi}$ is the third component of the orbital 
angular momentum and $\tau_3$ is the third
component of the isospin Pauli matrices.

We briefly explain the numerical method for the spectrum of this fermions. 
According to the Rayleigh-Ritz variational principle, the upper 
bound of the spectrum can be obtained from the secular equation for each 
${\cal K}$; 
\begin{eqnarray}
	\rm{det}\left({\mathcal A}-\epsilon {\mathcal B}\right) = 0 
	\label{secular_equation}
\end{eqnarray}
where 
\begin{eqnarray}
	{\mathcal A}_{k'^{(p)}k^{(q)}}&=&
	\int d^{3}x \phi_{\cal K}^{(p)} (k'^{(p)},\bm{x})^\dagger{\cal H}_{\rm fermi}
	\phi_{\cal K}^{(q)} (k^{(q)},\bm{x})\,, \nonumber \\
	{\mathcal B}_{k'^{(p)}k^{(q)}}&=&
	\int d^{3}x \phi_{\cal K}^{(p)} (k'^{(p)},\bm{x})^\dagger \phi_{\cal K}^{(q)} (k^{(q)},\bm{x})\,\, .
	\label{matrixelement}
\end{eqnarray}
The plain wave basis are defined as
\begin{eqnarray}
\phi^{(u)}_{\mathcal{K}}(k^{(u)}_i,\bm{x}):=
\langle \bm{x}|(u)\mathcal{K};i\rangle=
&\mathcal{N}^{(u)}_{i}
\left(
\begin{array}{c}
\omega^{(u)-}_{i,\epsilon} J_{\mathcal{K}-\frac{1}{2}-\frac{n}{2}}(k^{(u)}_i\rho)e^{i(\mathcal{K}-\frac{1}{2}-\frac{n}{2})\varphi}\\
\omega^{(u)+}_{i,\epsilon}  J_{\mathcal{K}+\frac{1}{2}-\frac{n}{2}}(k^{(u)}_i\rho)e^{i(\mathcal{K}+\frac{1}{2}-\frac{n}{2})\varphi}
\end{array}\right)
\otimes
\left(
\begin{array}{c}
1\\
0
\end{array}\right)\,,\label{eqm001}\\
\phi^{(d)}_{\mathcal{K}}(k_i^{(d)},\bm{x}):=
\langle \bm{x}|(d)\mathcal{K};i\rangle=
&\mathcal{N}^{(d)}_{i}
\left(
\begin{array}{c}
\omega^{(d)+}_{i,\epsilon}  J_{\mathcal{K}-\frac{1}{2}+\frac{n}{2}}(k^{(d)}_i\rho)e^{i(\mathcal{K}-\frac{1}{2}+\frac{n}{2})\varphi}\\
\omega^{(d)-}_{i,\epsilon}  J_{\mathcal{K}+\frac{1}{2}+\frac{n}{2}}(k^{(d)}_i\rho)e^{i(\mathcal{K}+\frac{1}{2}+\frac{n}{2})\varphi}
\end{array}\right)
\otimes
\left(
\begin{array}{c}
0\\
1
\end{array}\right)\,,\label{eqm002}
\end{eqnarray}
where
\begin{eqnarray}
\omega^{(p)+}_{i,\epsilon>0}=\omega^{(p)-}_{i,\epsilon<0}=1,~~
\omega^{(p)-}_{i,\epsilon>0}=\omega^{(p)+}_{i,\epsilon<0}=\frac{-{\rm sign}(\epsilon)k^{(p)}_i}{|\epsilon|+m},~~
p=u,d
\label{eqm003}
\end{eqnarray}
We can construct the plane wave basis as a cylinder of large radius $D$. 
As a result of imposing the boundary conditions
\begin{eqnarray}
J_{\mathcal{K}+\frac{1}{2}-\frac{n}{2}}(k^{(u)}_iD)=0,~~
J_{\mathcal{K}-\frac{1}{2}+\frac{n}{2}}(k^{(d)}_iD)=0
\label{eqm004}
\end{eqnarray}
we obtain a discrete set of wave numbers $k^{(u)}$ and $k^{(d)}$. 
We then have the orthogonality conditions
\begin{eqnarray}
&&\int^D_0d\rho \rho J_\nu(k^{(p)}_i\rho)J_\nu(k^{(p)}_j\rho)=\int^D_0d\rho \rho J_{\nu\pm 1}(k^{(p)}_i\rho)J_{\nu\pm 1}(k^{(p)}_j\rho)
=\delta_{ij}\frac{D^2}{2}[J_{\nu\pm 1}(k^{(p)}_iD)]^2
\end{eqnarray}
where $\nu=\mathcal{K}\pm\frac{1}{2}\mp\frac{n}{2}$. We can solve Eq.(\ref{secular_equation}) numerically. 
For the entire infinite set of wave numbers (which means an infinite size of the matrices), the eigenvalue 
becomes exact. The normalization constants of the basis vectors are
\begin{eqnarray}
\mathcal{N}^{(u)}_{i}=\left[
\frac{2\pi D^2|\epsilon|}{|\epsilon|+m}\left(J_{\mathcal{K}-\frac{1}{2}-\frac{n}{2}}(k^{(u)}_iD)\right)^2
\right]^{-\frac{1}{2}},~~
\mathcal{N}^{(d)}_{i}=\left[
\frac{2\pi D^2|\epsilon|}{|\epsilon|+m}\left(J_{\mathcal{K}+\frac{1}{2}+\frac{n}{2}}(k^{(d)}_iD)\right)^2
\right]^{-\frac{1}{2}}\,.
\label{eqm005}
\end{eqnarray}

To obtain the spectral flow, we use the linearly interpolated field
\begin{eqnarray}
U_{\rm intp}(x,\lambda)=(1-\lambda)U_\infty+\lambda U(x),~~~~0\leq \lambda \leq 1
\label{intpfield}
\end{eqnarray}
where $U(x)$ is the field with a nontrivial topology and $U_\infty$ is the asymptotic field. For a given $\lambda$, 
we smoothly connect the vacuum and the solitonic states. We substitute $U_{\rm intp}(x,\lambda)$ in Hamiltonian  (\ref{hamiltonian_bs}) 
and solve eigenproblem  (\ref{secular_equation}) by the standard matrix diagonalization algorithm of LAPACK.

In Fig.\ref{spectralflow_bs}, we present some typical spectral flow results. Some special energy levels pass from the
positive to the negative continuum. The number of fermionic levels crossing zero is always equal to the
topological charges of $U(x)$. As is seen, these levels are normalizable modes, and the behavior is then
described by the index theorem.

The eigenfunction $\psi(\bm{x})$ of Hamiltonian  (\ref{hamiltonian_bs}) in terms of the eigenstates of (\ref{secular_equation}) has the form
\begin{eqnarray}
\psi_{\mathcal{K},\mu}(\bm{x}):=\langle \bm{x}|\mathcal{K};\mu \rangle
=\sum_{i}\Bigl[
\langle \bm{x}|(u)\mathcal{K};i\rangle \langle (u)\mathcal{K};i|\mu\rangle
+\langle \bm{x}|(d)\mathcal{K};i\rangle \langle (d)\mathcal{K};i|\mu\rangle
\Bigr]\,.
\label{eigenfunction}
\end{eqnarray}
We note that the grand spin $\mathcal{K}$ is a good quantum number; matrices (\ref{matrixelement}) are block diagonal. 
Therefore, the summation for each $\mathcal{K}$ in (\ref{eigenfunction}) is only over $i$. 
Hence, we can compute the normalized density of the fermion mode
\begin{eqnarray}
d_\mathcal{K}(\rho)=\frac{1}{2\pi n_c}\int d\varphi \bar{\psi}_{\mathcal{K},\mu=0}(\bm{x})\gamma_3\psi_{\mathcal{K},\mu=0} (\bm{x}) .
\label{density}
\end{eqnarray}
where $\psi_{\mathcal{K},\mu=0}$ indicates the mode of the spectral flow. Plots of the density are shown in Fig.\ref{fermiondensity_bs}. 
For numerical calculations, we chose the cylinder radius $D = 50.0$ and $512$ points for discretizing the momenta. 
These spectral flow modes localize on the soliton and then become normalizable modes. The results show that the
anyon angle $\Theta\equiv n_c\pi {\rm sgn}(m)$ is determined by the number of the normalizable fermionic modes. The nature
of the spin statistics of the baby skyrmion is thus consistent with the nature of the localized fermions.

\section{The $\mathbb{C}P^N$ model}
\subsection{The model and the field equations}
We briefly sketch the Skyrme-Faddeev model on the target space $\CP^N$. We introduce 
the Lagrangian of the form
\begin{eqnarray}
{\cal L}_{\rm SF}&=&\frac{M^2}{2}{\rm Tr}(\partial_\mu\Phi\partial^\mu\Phi)
+\frac{1}{e^2}{\rm Tr}([\partial_\mu\Phi,\partial_\nu\Phi]^2) \nonumber \\
&+&\frac{\beta}{2}\Bigl({\rm Tr}(\partial_\mu\Phi\partial^\mu\Phi)\Bigr)^2
+\gamma\Bigl({\rm Tr}(\partial_\mu\Phi\partial_\nu\Phi)\Bigr)^2
-\mu^2V(\Phi)
\label{actionp}
\end{eqnarray}
where $M^2$ is a coupling constant with the dimension of mass, the coupling constants $e^{?2}$, $\beta$, and $\gamma$ 
have the dimension of inverse mass, and $V$ denotes a potential term that contains no derivative term and does not
break local symmetries of the model. For the field $\Phi$, we use the parameterization with $N$ complex fields
$u:=\{u_i\},~~i=1,\ldots,N$,
\begin{eqnarray}
\Phi=
I_{N+1\times N+1}
+
\frac{2}{\vartheta^2}\biggl(\begin{array}{cc}
-u\otimes u^\dagger & iu \nonumber \\
-iu^\dagger & -1 \nonumber 
\end{array}\biggr),~~~~
\vartheta:=\sqrt{1+u^\dagger\cdot u}\,.
\end{eqnarray}
In terms of those fields, 
the Lagrangian (\ref{actionp}) is written as
\begin{eqnarray}
{\cal L}_{\rm SF}=
-\frac{1}{2} \Bigl[M^2 \eta_{\mu\nu}+C_{\mu\nu}\Bigr]\tau^{\nu\mu}-\mu^2V
\label{actioncmunu}
\end{eqnarray}
where 
\begin{eqnarray}
&&C_{\mu\nu}:= M^2\eta_{\mu\nu}-\frac{4}{e^2}\Bigl[(\beta e^2-1)\tau_{\rho}^{\rho}\eta_{\mu\nu}
+(\gamma e^2-1)\tau_{\mu\nu}+(\gamma e^2+2)\tau_{\nu\mu}\Bigr],\\
\label{cmunudef}
&&\tau_{\mu\nu}:=-\frac{4}{\vartheta^4}\left[\vartheta^2
\partial_{\nu}u^{\dagger}\cdot \partial_{\mu}u-(\partial_{\nu}u^{\dagger}\cdot u)(u^{\dagger}\cdot\partial_{\mu}u) \right].
\end{eqnarray}
Variation with respect to $u_i^*$ leads to equations that after multiplication by the function inverse to $\Delta^2_{ij}$
, i.e., by
\begin{align}
\Delta^{-2}_{ij}=\frac{1}{1+u^{\dagger}\cdot u}(\delta_{ij}+u_i u^{*}_j),\nonumber 
\end{align}
can be written in the form
\begin{eqnarray}
&&(1+u^{\dagger}\cdot u)\partial^{\mu}(C_{\mu\nu}\partial^{\nu}u_i)
-C_{\mu\nu}\left[(u^{\dagger}\cdot\partial^{\mu}u)\partial^{\nu}u_i+(u^{\dagger}\cdot\partial^{\nu}u)\partial^{\mu}u_i\right]\nonumber\\
&&\hspace{3cm}+\frac{\mu^2}{4}(1+u^{\dagger}\cdot u)^2\sum_{k=1}^{N}\left[(\delta_{ik}+u_iu^*_k)\frac{\delta V}{\delta u^*_k}\right]=0\,.
\label{eom1}
\end{eqnarray}
In what follows, we consider some example potentials. In the simplest case, where the potential is a function
of absolute values of the fields, $V(|u_1|^2,\ldots,|u_N|^2)$, the contribution related to this potential becomes
$$
\sum_{k=1}^{N}\left[(\delta_{ik}+u_iu^*_k)\frac{\delta V}{\delta u^*_k}\right]
=u_i\left[\frac{\delta V}{\delta |u_i|^2}+\sum_{k=1}^N|u_k|^2\frac{\delta V}{\delta |u_k|^2}\right].
$$

We consider the ansatz
\begin{eqnarray}
u_j=f_j(\rho)e^{in_j\varphi}\,.\label{ansatz}
\end{eqnarray}
The constants ni form a set of integers. We define the diagonal matrix $\lambda\equiv{\rm diag}(n_1,\ldots,n_N)$
to simplify the form of some formulas below. The expressions $\tau_{\mu\nu}$ have the forms
\begin{eqnarray}
&&\tau_{\rho\rho}\equiv\theta(\rho)=-\frac{4}{\vartheta^4}\,\left[\vartheta^2\,f'^T.f'-(f'^T.f)(f^T.f')\right]\,,\nonumber\\
&&\tau_{\varphi\varphi}\equiv\omega(\rho)=-\frac{4}{\vartheta^4}\,\left[\vartheta^2\,f^T.\lambda^2.f-(f^T.\lambda.f)^2\right]\,,\nonumber\\
&&\tau_{\varphi\rho}=-\tau_{\rho\varphi}\equiv i\zeta(\rho)=
-i\frac{4}{\vartheta^4}\,\left[\vartheta^2\,f'^T.\lambda.f-(f^T.\lambda.f)(f'^T.f)\right] 
\end{eqnarray}
where the prime denotes the derivative with respect to $\rho$ and $T$ denotes transposition. The equations of
motion in dimensionless coordinates become
\begin{eqnarray}
&&(1+f^T.f)\left[\frac{1}{\rho}\left(\rho\,\tilde{C}_{\rho\rho}f'_k\right)'+\frac{i}{\rho}\left(\frac{\tilde{C}_{\rho\varphi}}{\rho}\right)'(\lambda.f)_k-\frac{1}{\rho^4}\tilde{C}_{\varphi\varphi}(\lambda^2.f)_k\right]\nonumber\\
&&-2\left[\tilde{C}_{\rho\rho}(f^T.f')f'_k-\frac{1}{\rho^4}\tilde{C}_{\varphi\varphi}(f^T.\lambda.f)(\lambda.f)_k\right]
+\tilde{\mu}^2\frac{f_k}{4}(1+f^{T}.f)^2\left[\frac{\delta V}{\delta f_k^2}+\sum_{i=1}^Nf_i^2\frac{\delta V}{\delta f_i^2}\right]=0~~~~
\label{equationr}
\end{eqnarray}
for each $k=1,\ldots,N$, where we have introduced the symbols 
$\tilde {C}_{\mu\nu}:= \frac{1}{M^2}C_{\mu\nu}$, and also $\tilde{\mu}^2:=\frac{r_0^2}{M^2}\mu^2$. 
The components $\tilde{C}_{\mu\nu}$ in the equations of motion are
\begin{eqnarray}
&&\tilde{C}_{\rho\rho}=-1+(\beta e^2-1)\left(\theta+\frac{\omega}{\rho^2}\right)+(2\gamma e^2+1)\theta\,,\nonumber\\
&&\tilde{C}_{\varphi\varphi}=-\rho^2+\rho^2(\beta e^2-1)\left(\theta+\frac{\omega}{\rho^2}\right)+(2\gamma e^2+1)\omega\,,\nonumber\\
&&\tilde{C}_{\varphi\rho}=-\tilde{C}_{\rho\varphi}=-3i\zeta\,.
\end{eqnarray}
In the numerical computation, it is useful to introduce the scaled coordinate y and the variables $g_j$ :
\begin{eqnarray}
\rho=\sqrt{\frac{1-y}{y}},~~f_j=\frac{1}{\sqrt{N}}\sqrt{\frac{1-g_j}{g_j}},~~~~y\in(0,1],~~~~ g_j\in(0,1].
\end{eqnarray}

Taking the results in \cite{DAdda:1980zni} and also the discussion in \cite{Ferreira:2010jb} into account, 
we can determine the topological charge in the present model. 
The field $u_i$ yields a map from the plane $(x^1, x^2)$ to $\mathbb{C}P^N$. But for the energy
to be finite, the field must tend to a constant at spatial infinity. The plane $(x^1, x^2)$ is then topologically
compactified into $S^2$, and the finite energy field configuration defines a map $S^2\to \CP^N$, which is classified
into the homotopy classes of the group $\pi_2(\CP^N)$. There is a theorem \cite{DAdda:1980zni} 
according to which $\pi_2(G/H)=\pi_1(H)_G$ where $\pi_1(H)_G$ is the subset of $\pi_1(H)$ formed by closed paths 
in $H$ that can be contracted to a point in $G$. In our case, the homotopy group is thus given by
\begin{eqnarray}
\pi_2(\CP^N)=\pi_1(SU(N)\otimes U(1))_{SU(N+1)}=\integer.
\end{eqnarray}
The topological charge, an element of the homotopy group, is given by the integral of the 
 topological current defined in terms of the field $\Phi$ as
\begin{eqnarray}
j^\mu (\Phi)=\frac{i}{16\pi}\epsilon^{\mu\nu\lambda}{\rm Tr}(\Phi\partial_\nu \Phi\partial_\lambda \Phi).
\label{topologicalcurrent}
\end{eqnarray}
As noted in~\cite{Ferreira:2010jb,DAdda:1978vbw}, the topological charge is in fact equal to the number of poles of $u_i$ 
including poles at infinity. Because the solutions behave as a holomorphic function near the boundaries, i.e., $u_i\sim\rho^{n_i}e^{i n_i\varphi}$
(where $n_i \in \mathrm{Z}$) near the origin and at spatial infinity, the topological charge is given by $Q_{\rm top}=n_{\rm max}+|n_{\rm min}|$,
where $n_{\rm max}$ and $n_{\rm min}$ 
are the greatest positive integer and the least negative integer in the set $n_1,\cdots,n_N$.

We now give the potential term in explicit form. In the general case, potential terms are a function
of fields, which vanish at spatial infinity and preserve the local symmetries of the model. In this model,
the simplest choice is the gold babyh-type potential $\mathrm{Tr}(1-\Phi^{-1}_\infty\Phi)$, where $\Phi_\infty$ 
is the value of the field at spatial infinity, i.e., $\Phi_{\infty}:=\lim_{\rho\to\infty}\Phi(\rho)$. 
Assuming that the solution and its holomorphic counterpart have the same asymptotic behavior at spatial 
infinity, we find that inverse of the principal variable $\Phi$ goes to
${\Phi_{\infty}}^{-1}:={\rm diag}(-1,1,1)$ as $\rho\to\infty$ for $n_1 > n_2 > 0$. 
We note that the inverse of the principal variable goes
to $\Phi_0^{-1}:={\rm diag}(1,1,-1)$ as $\rho \to 0$. 
The expression ${\rm Tr}(1-\Phi_0^{-1}\Phi)$ can then be included as the gnew babyh
potential with two vacuums~\cite{Kudryavtsev:1997nw}. Finally, we consider the general form of the potential
\begin{eqnarray}
V&=&[{\rm Tr}(1-\Phi_0^{-1}\Phi)]^a [{\rm Tr}(1-\Phi_\infty^{-1} \Phi)]^b \nonumber \\
&=&\frac{(|u_1|^2+|u_2|^2)^a(1+|u_2|^2)^b}{(1+|u_1|^2+|u_2|^2)^{a+b}}
=\frac{(g_1+g_2-2g_1g_2)^ag_1^b(1+g_2)^b}{(g_1+g_2)^{a+b}}
\label{potcp2++}
\end{eqnarray}
where the integers $a\geq 0$ and $b>0$. 

Assuming that for $n_2 < 0$, the field $u_2$ behaves at zero as its holomorphic counterpart, i.e., $\sim\rho^{n_2}$ , 
we find that it tends to diverge as $\rho\to 0$. 
The inverse of the principal variable $\Phi$ then goes to $\Phi_0^{-1}:={\rm diag}(1,-1,1)$
as $\rho\to 0$. The general form of the potential becomes
\begin{eqnarray}
V&=&\frac{(1+|u_1|^2)^a(1+|u_2|)^b}{(1+|u_1|^2+|u_2|^2)^{a+b}}
=\frac{g_1^bg_2^a(1+g_1)^a(1+g_2)^b}{(g_1+g_2)^{a+b}}
\label{potcp2+-}
\end{eqnarray}
where the integers satisfy $a\geq 0$ and $b>0$. 

\begin{figure*}[t]
  \begin{center}
	\hspace{-1.0cm}
    	\includegraphics[width=60mm]{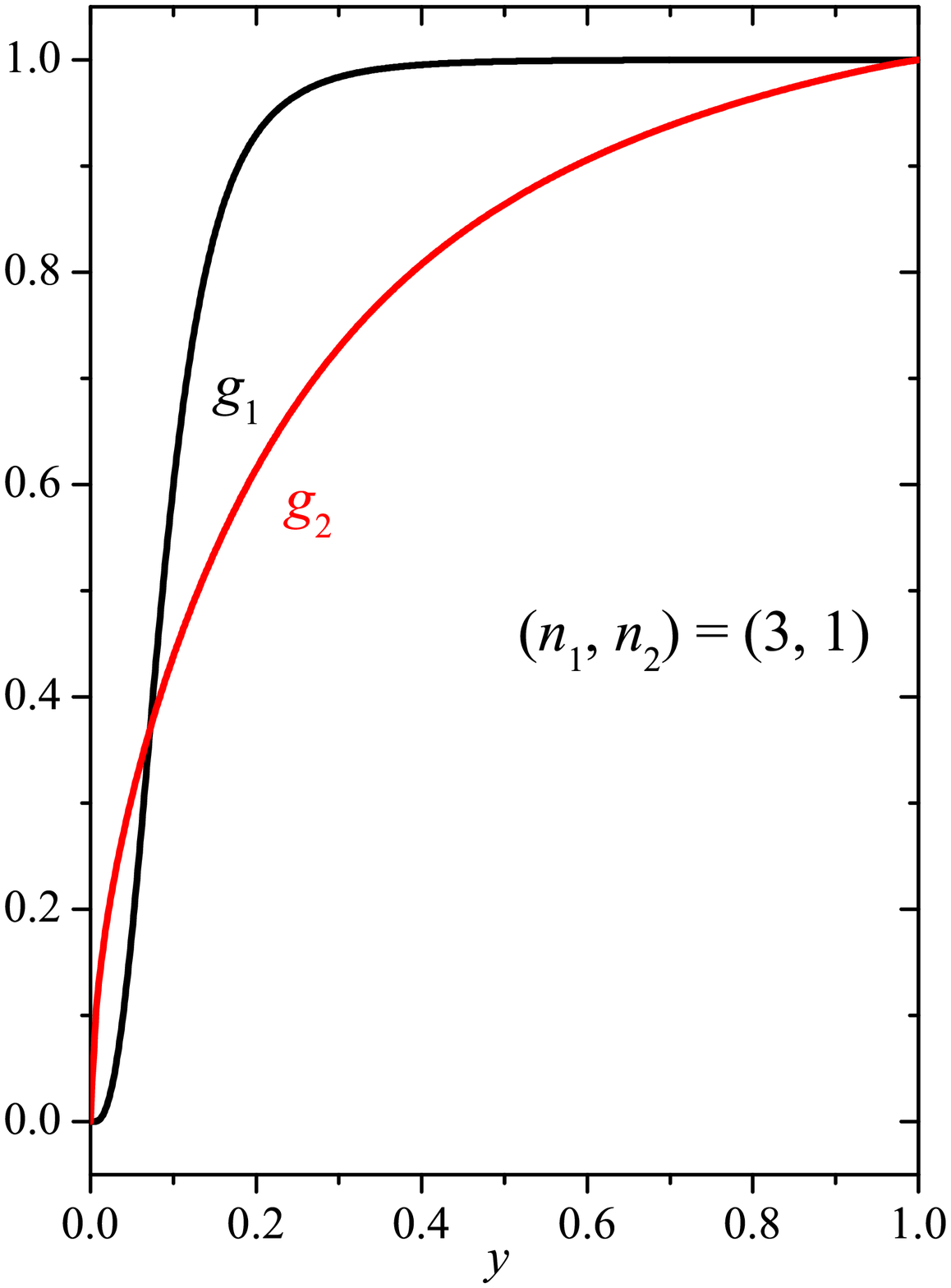}\hspace{-1.0cm}
	\includegraphics[width=60mm]{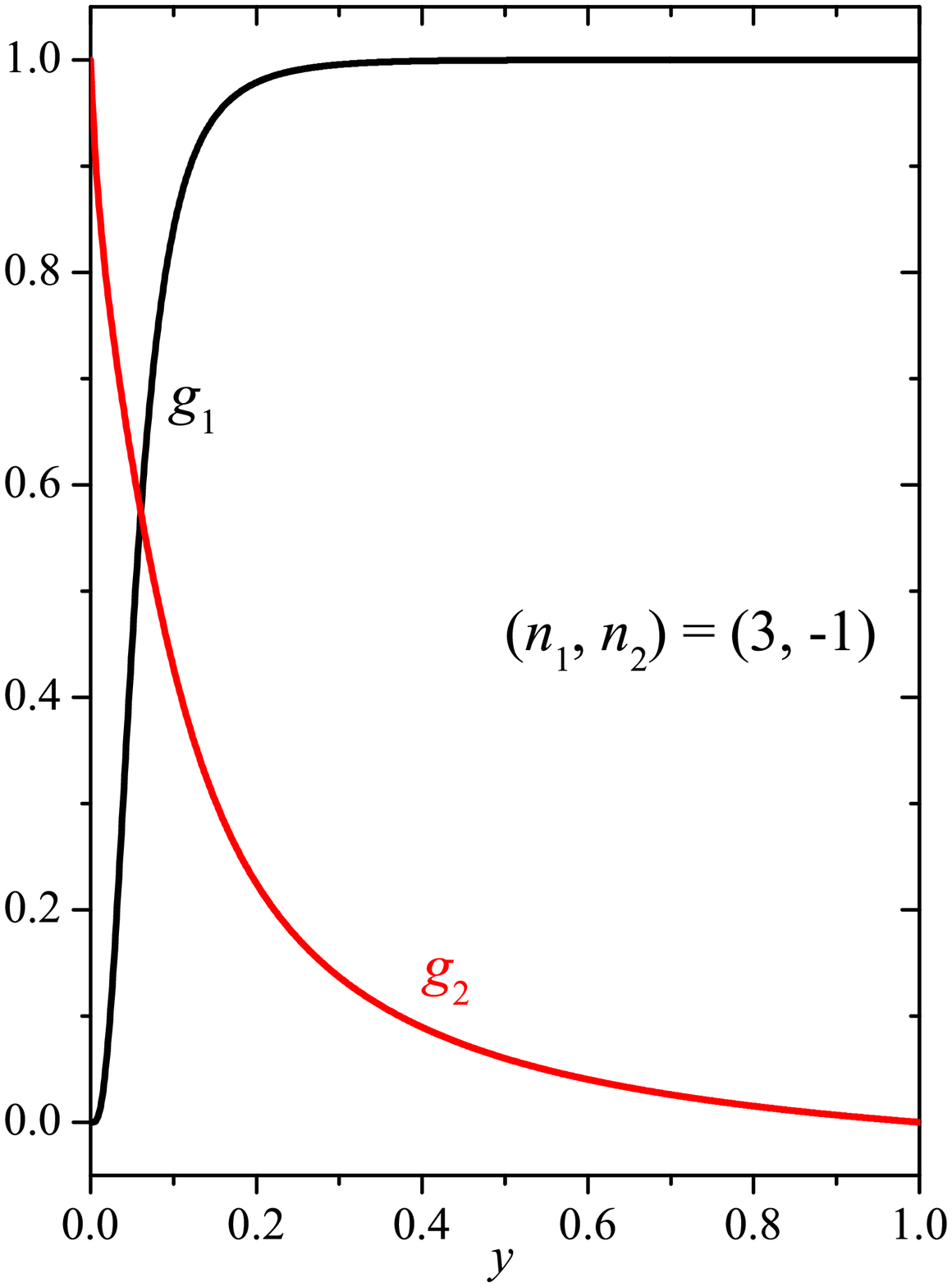}\hspace{-1.0cm}
 	\includegraphics[width=60mm]{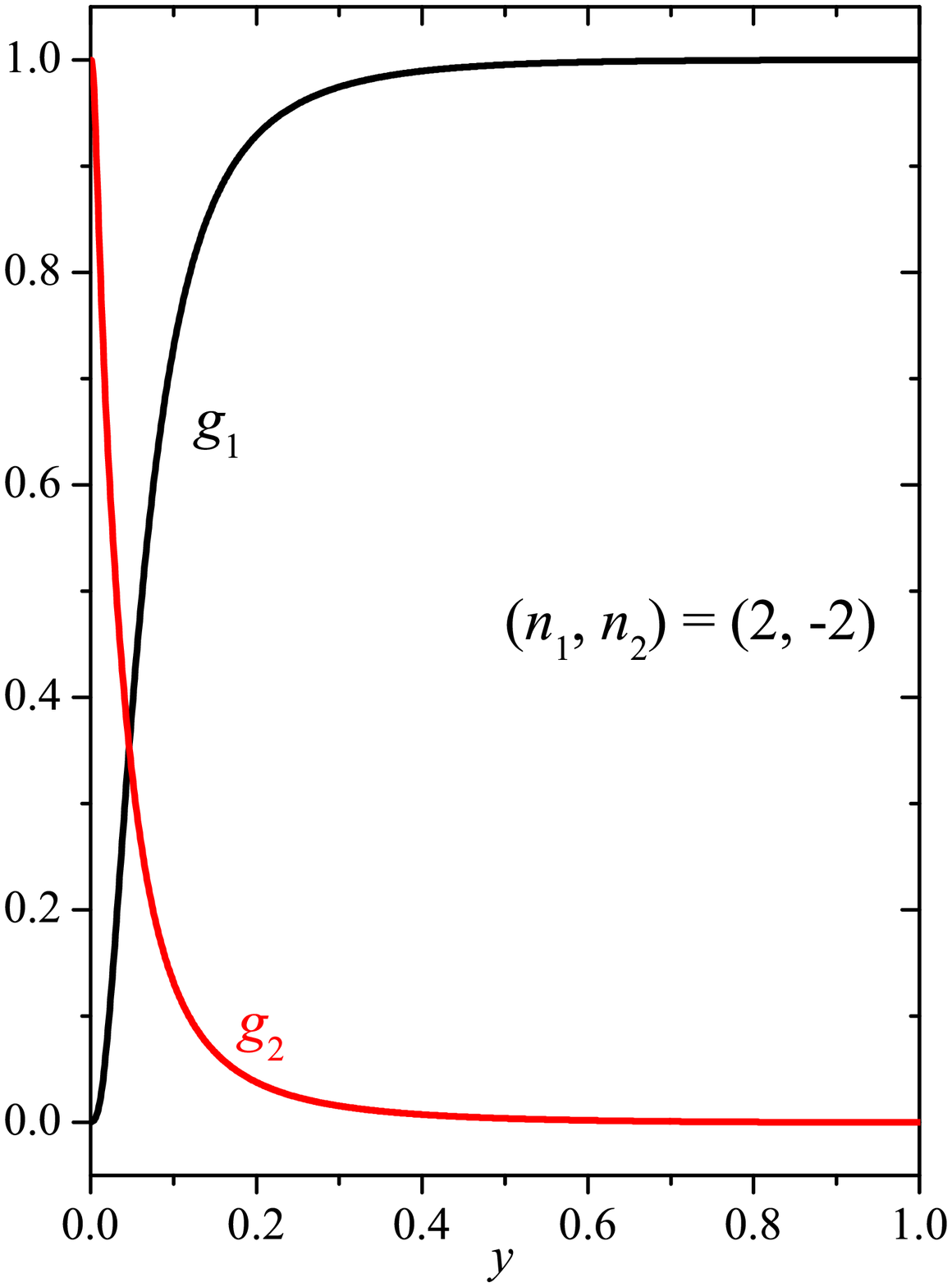}
    \caption{\label{profiles}The plot of the solutions $g_1,g_2$ with the potential $(a,b)=(0,2)$.
The other parameters are $(\beta e^2,\gamma e^2,\tilde{\mu}^2)=(5.0,5.0,1.0)$.}
  \end{center}
	\end{figure*}

\subsection{Normalizable modes of fermions}

For the target space $\mathbb{C}P^N$, fermions with chiral symmetry
coupled to the soliton were first discussed in \cite{DAdda:1980zni}. It was confirmed that the normalizable zero mode of the
fermion appears by virtue of the index theorem.

We consider a gauged model corresponding to the Lagrangian (\ref{actionp})
\begin{eqnarray}
{\cal L}_{\rm fermi}=\bar{\psi}i\gamma^\alpha(\partial_\alpha-iA_\alpha)\psi-m\bar{\psi}\Phi\psi
\equiv \bar{\psi}i{\cal D}_A\psi
\end{eqnarray}
where $\alpha=1,2,3$. The gamma matrices are the standard prescription such that 
$\gamma^1=-i\sigma_1, \gamma^2=-i\sigma_2, \gamma^3=\sigma_3$ where $\sigma_\alpha$ are standard Pauli matrices. 

The Euclidean partition function is defined as
\begin{eqnarray}
{\cal Z}=e^{\omega(\Phi)}=\int{\cal D}\psi{\cal D}\bar{\psi}\exp(\bar{\psi}i{\cal D}_A\psi)=(\det i{\cal D}_A)^{n_c}.
\end{eqnarray}
We obtain the real and the imaginary part of the effective action:
\begin{eqnarray}
&&\omega(\Phi)_{\rm Re}|_{A=0}=n_c\frac{|m|}{16\pi}\int d^3x{\rm Tr}(\partial_\mu \Phi\partial^{\mu} \Phi)
+({\rm higher~derivative~terms}),
\label{effectiveactioncpn_fr}
\\
&&\omega(\Phi)_{\rm Im}=-n_c\int d^3x\Bigl(j^\mu (\Phi) A_\mu+\pi\,{\rm sgn}(m){\cal L}_{\rm Hopf}(\Phi)\Bigr)\,
\label{effectiveactioncpn_fi}
\end{eqnarray}
where $n_c$ is a degeneracy of the fermions. 
The explicit form of the current $j^\mu$ coincides with (\ref{topologicalcurrent}). Consequently, as noted 
in~\cite{Jaroszewicz:1985ip,Abanov:2000ea,Abanov:2001iz}, the anyon
angle $\Theta$ is determinable in this fermionic context as $\Theta\equiv n_c\pi\,{\rm sgn}(m)$ if the vortices are coupled to the
fermionic field. But because $\Pi_3(\mathbb{C}P^N)$ is trivial, the Hopf term ${\cal L}_{\rm Hopf}$ itself is perturbative, and the value
of the integral depends on the background classical solutions. Consequently, we cannot expect that this
value becomes an integer. As a result, the solitons are always anyons even if $\Theta=n\pi, n\in \mathbb{Z}$~\cite{Amari:2016ynl}.

\begin{figure*}[t]
  \begin{center}
    	\includegraphics[width=110mm]{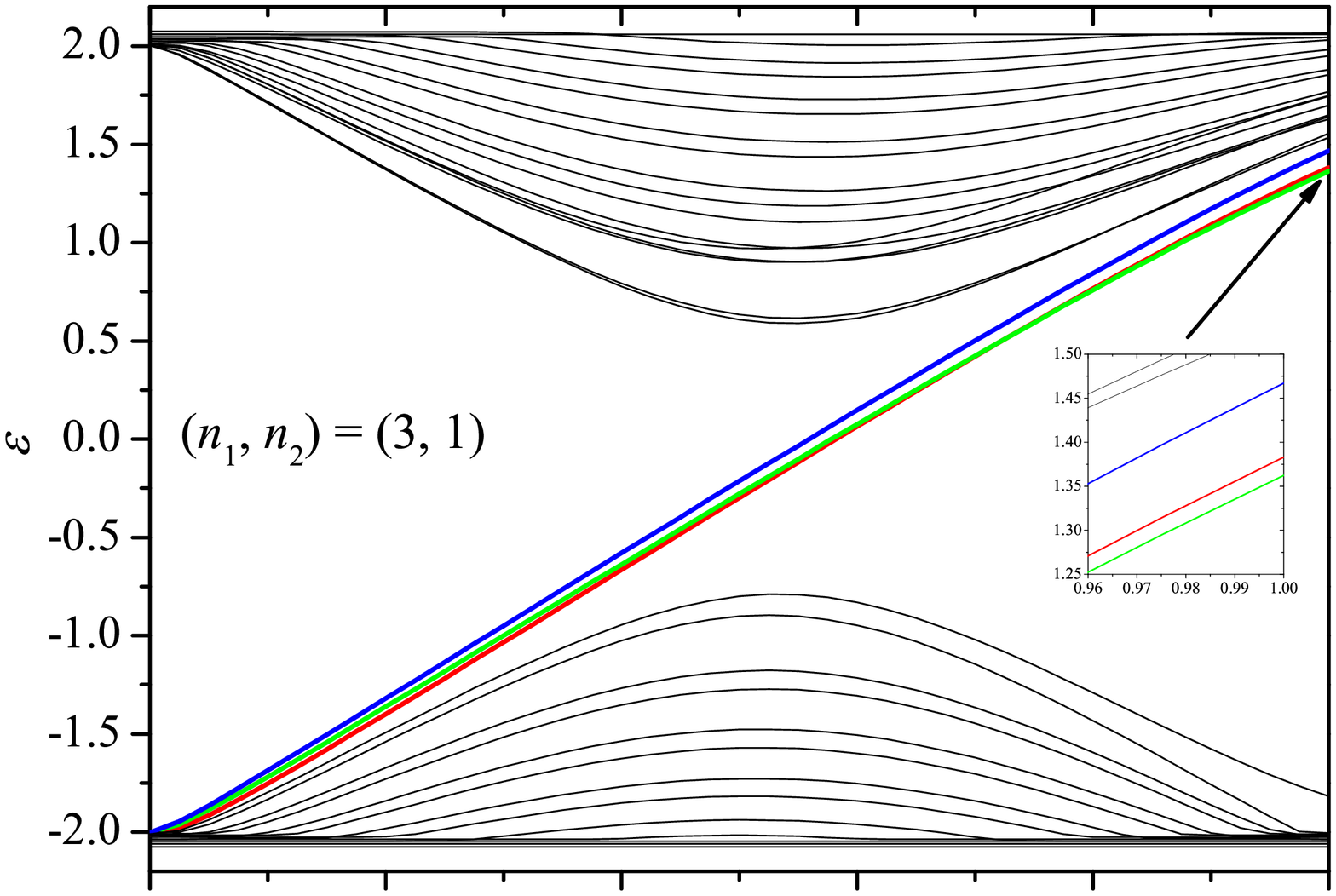}
\vspace{-1cm}

 	\includegraphics[width=110mm]{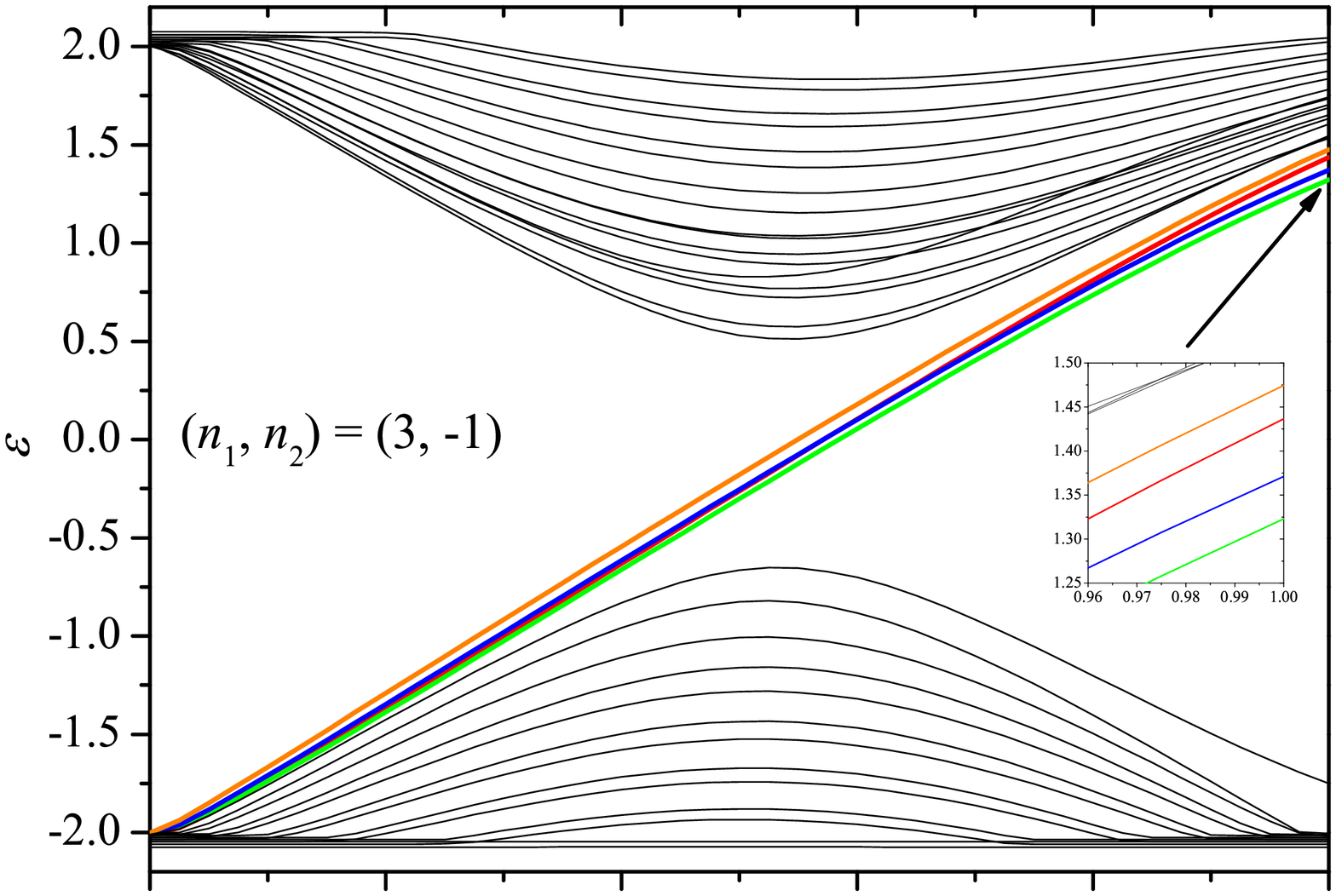}
\vspace{-1cm}

 	\includegraphics[width=110mm]{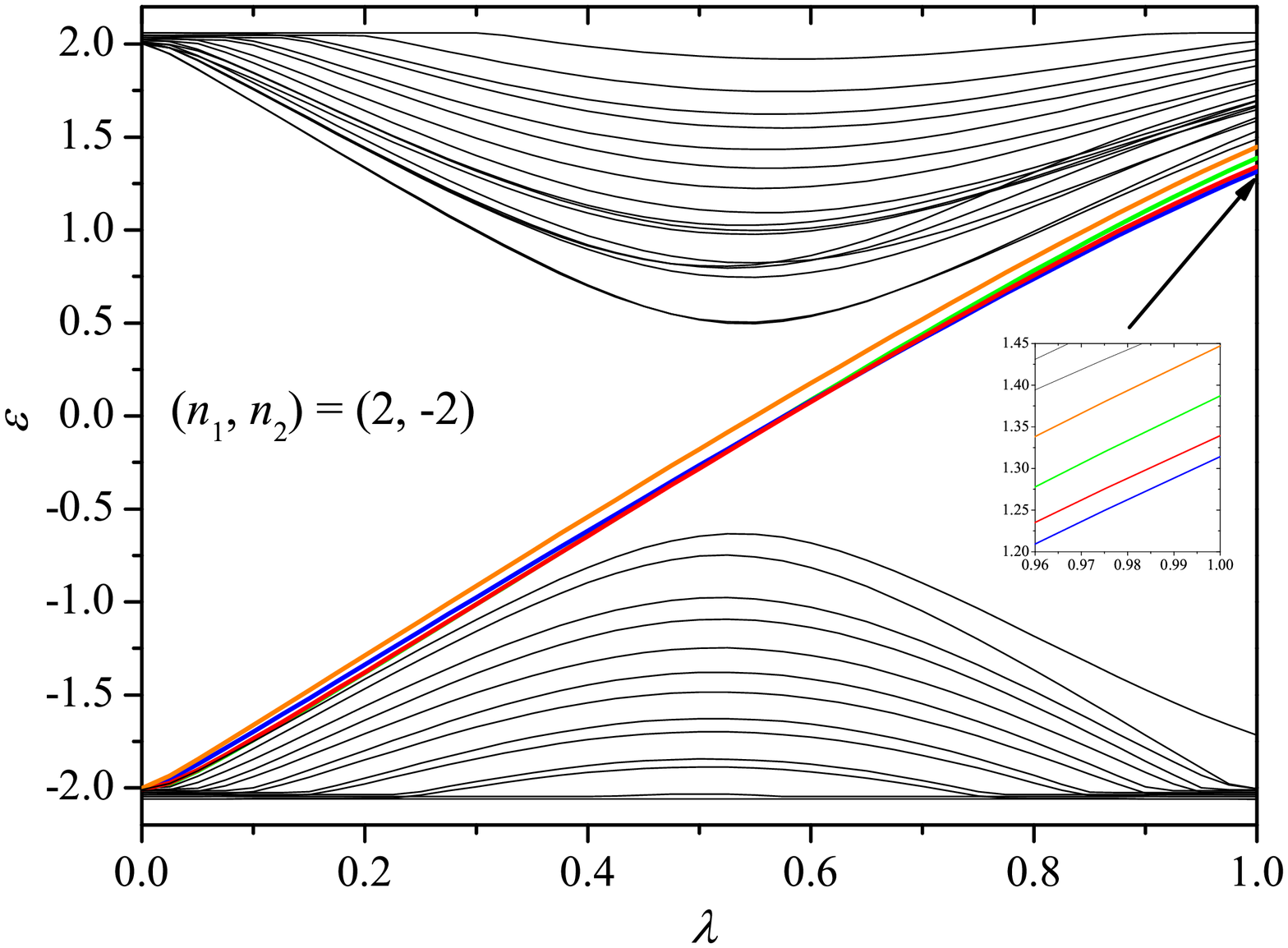}

    \caption{\label{spectralflow} 
The spectral flow corresponding to the solutions in Fig. \ref{profiles}: the first 34 levels (17 positive
and 17 negative for the vacuum background $\lambda=0$) are shown. The coupling constant is chosen as
$m = 2.0$.
}
  \end{center}
	\end{figure*}

\begin{figure*}[t]
  \begin{center}
    	\includegraphics[width=110mm]{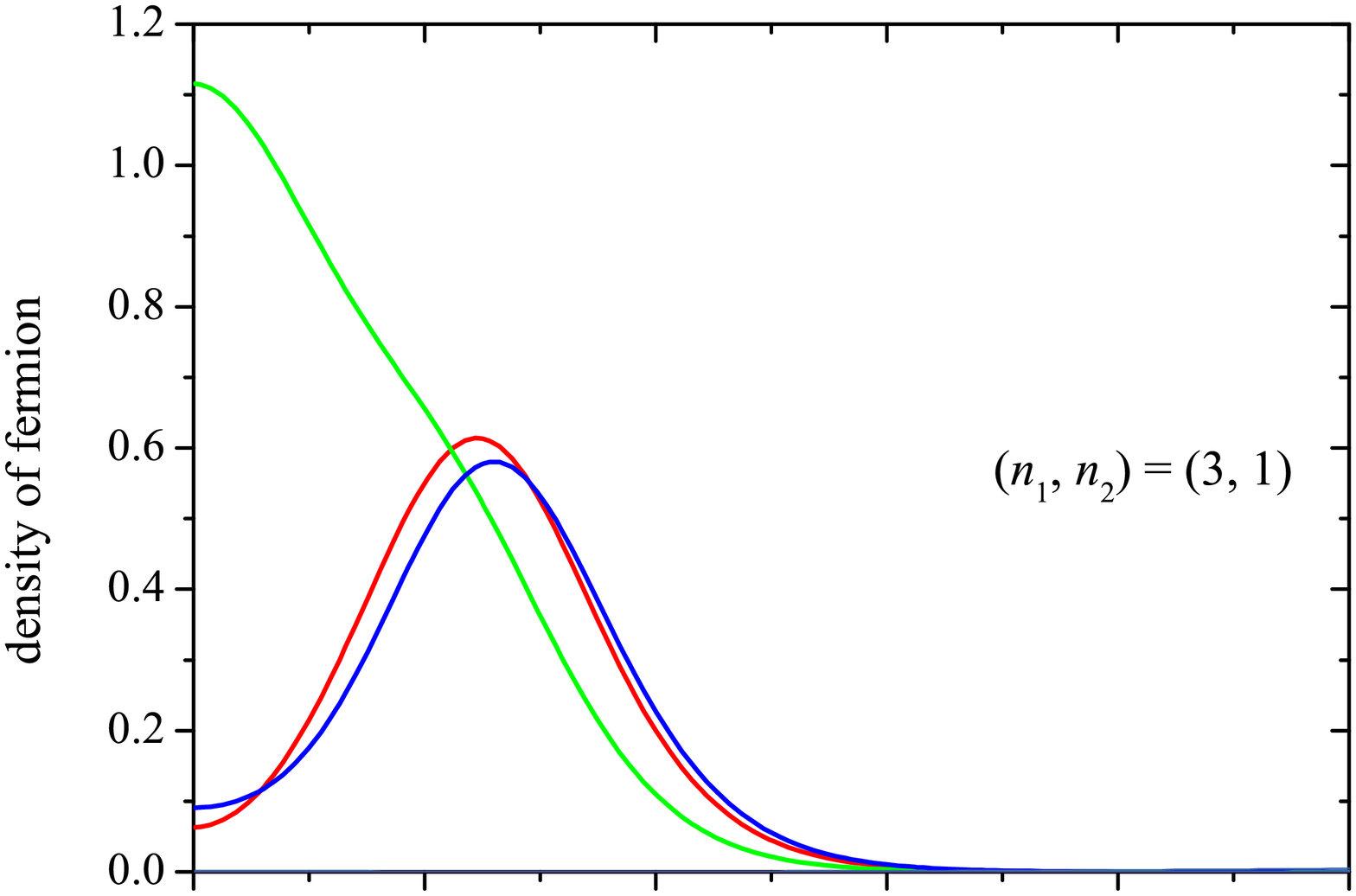}\\
\vspace{-1.0cm}

	\includegraphics[width=110mm]{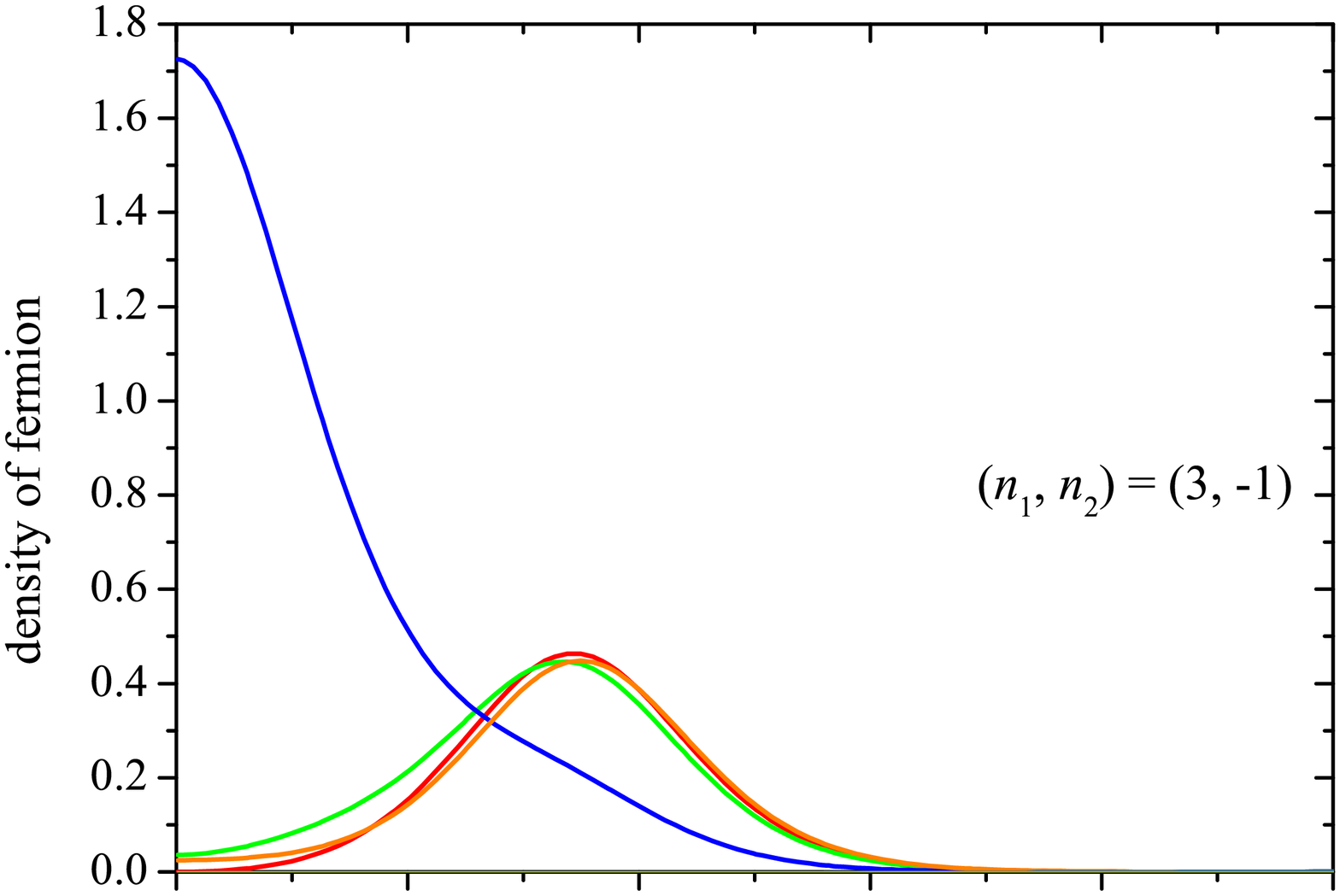}\\
\vspace{-1.0cm}
	
 	\includegraphics[width=110mm]{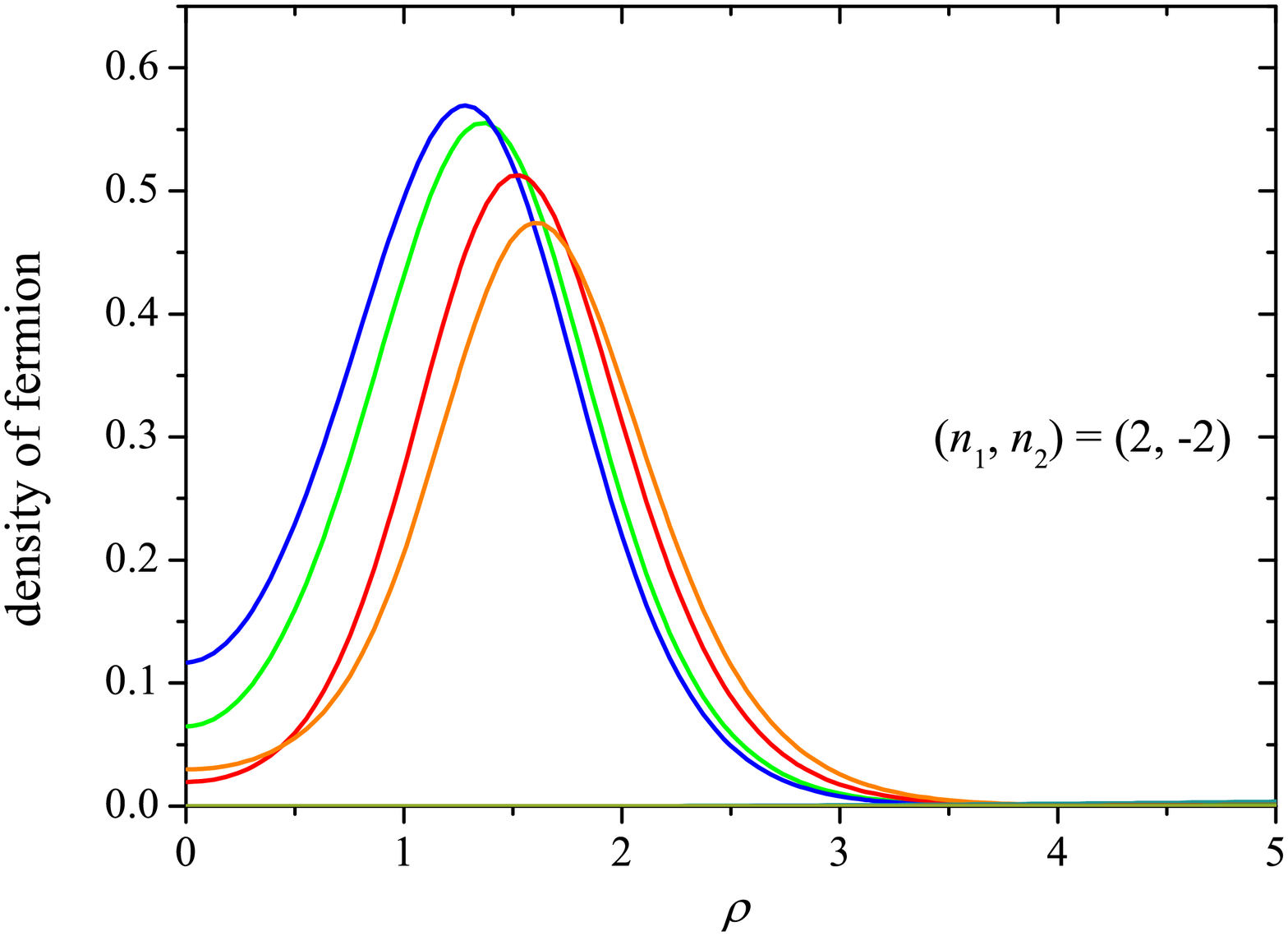}
    \caption{\label{fermiondensity}
Fermion density (\ref{density}) for the model with potential (\ref{potcp2++}) with $(a,b)=(0,2)$, and the other
parameters $(\beta e^2,\gamma e^2,\tilde{\mu}^2)=(1.1,1.1,1.0)$: the eigenfunctions are computed with the solitonic background
at $\lambda=1$. The curve types correspond to the curves in the insets in Fig. \ref{spectralflow}.
}
  \end{center}
	\end{figure*}

\subsection{The numerical analysis}

In Fig.\ref{profiles}, we show plots of typical soliton solutions for the topological
charges $Q_{\rm top} = 3, 4$. 
We chose the potential with $a = 0$ and $b = 2$ in  (\ref{potcp2++}), or (\ref{potcp2+-}).

The Hamiltonian has the form
\begin{eqnarray}
{\cal H}_{\rm fermi}=-i\gamma_3\gamma_k\partial_k+\gamma_3m\Phi
=
\left(
\begin{array}{cc}
m\Phi & -e^{-i\varphi}\Bigl(\partial_\rho-\frac{i\partial_\varphi}{\rho}\Bigr) \\
e^{i\varphi}\Bigl(\partial_\rho+\frac{i\partial_\varphi}{\rho}\Bigr)& -m\Phi  \\
\end{array}
\right)\,.
\label{hamiltonian}
\end{eqnarray}
It can be shown that ${\cal H}_{\rm fermi}$ commutes with the operator of angular
momentum, which we call {\it a grand spin} as before,
\begin{eqnarray}
{\cal K}=\ell_3+\frac{\sigma_3}{2}-\frac{n_1}{2}\biggl(\lambda_3+\frac{1}{\sqrt{3}}\lambda_8\biggr)
+\frac{n_2}{2}\biggl(\lambda_3-\frac{1}{\sqrt{3}}\lambda_8\biggr)
\label{conserv2}
\end{eqnarray}
where $\ell_3=({\bf r}\times{\bf p})_3=-i\frac{\partial}{\partial\varphi}$ and $\lambda_3,\lambda_8$ 
are Gell-Mann matrices.

The plain wave basis is
\begin{eqnarray}
&&\phi_{\cal K}^{(u)}(k_i^{(u)},\bm{x})={\cal N}^{(u)}_i
\left( 
	\begin{array}{c}
	\omega_{i,\epsilon}^{(u)-}J_{{\cal K}-\frac{1}{2}+\frac{2n_1-n_2}{3}}(k_i^{(u)}\rho)e^{i({\cal K}-\frac{1}{2}+\frac{2n_1-n_2}{3})\varphi}  
	\\
	\omega_{i,\epsilon}^{(u)+}J_{{\cal K}+\frac{1}{2}+\frac{2n_1-n_2}{3}}(k_i^{(u)}\rho)e^{i({\cal K}+\frac{1}{2}+\frac{2n_1-n_2}{3})\varphi}
	\end{array}
\right)
	\otimes
\left(
	\begin{array}{c}
	1 \\0 \\0
	\end{array}
\right)\,,	\nonumber \\
&&\phi_{\cal K}^{(d)}(k_i^{(d)},\bm{x})={\cal N}^{(d)}_i
\left( 
	\begin{array}{c}
	\omega_{i,\epsilon}^{(d)+}J_{{\cal K}-\frac{1}{2}-\frac{n_1-2n_2}{3}}(k_i^{(d)}\rho)e^{i({\cal K}-\frac{1}{2}-\frac{n_1-2n_2}{3})\varphi}  
	\\
	\omega_{i,\epsilon}^{(d)-}J_{{\cal K}+\frac{1}{2}-\frac{n_1-2n_2}{3}}(k_i^{(d)}\rho)e^{i({\cal K}+\frac{1}{2}-\frac{n_1-2n_2}{3})\varphi}
	\end{array}
\right)
	\otimes
\left(
	\begin{array}{c}
	0 \\1 \\0
	\end{array}
\right)\,,	\\
&&\phi_{\cal K}^{(s)}(k_i^{(s)},\bm{x})={\cal N}^{(s)}_i
\left( 
	\begin{array}{c}
	\omega_{i,\epsilon}^{(s)+}J_{{\cal K}-\frac{1}{2}-\frac{n_1+n_2}{3}}(k_i^{(s)}\rho)e^{i({\cal K}-\frac{1}{2}-\frac{n_1+n_2}{3})\varphi}  
	\\
	\omega_{i,\epsilon}^{(s)-}J_{{\cal K}+\frac{1}{2}-\frac{n_1+n_2}{3}}(k_i^{(s)}\rho)e^{i({\cal K}+\frac{1}{2}-\frac{n_1+n_2}{3})\varphi}
	\end{array}
\right)
	\otimes
\left(
	\begin{array}{c}
	0 \\0 \\1
	\end{array}
\right)	\nonumber 
\end{eqnarray}
where 
\begin{eqnarray}
\omega_{i,\epsilon_>0}^{(p)+}=\omega_{i,\epsilon<0}^{(p)-}=1,~~~~
\omega_{i ,\epsilon>0}^{(p)-}=\omega_{i,\epsilon<0}^{(p)+}=\frac{-{\rm sgn}(\epsilon)k_i^{(p)}}{|\epsilon|+m},~~~~p=u,d,s
\end{eqnarray}
We construct the plane wave basis in a cylinder of large radius $\rho=D$. 
As a result of imposing the boundary conditions
\begin{eqnarray}
J_{{\cal K}-\frac{1}{2}+\frac{2n_1-n_2}{3}}(k^{(u)}_iD)=0,~~
J_{{\cal K}-\frac{1}{2}-\frac{n_1-2n_2}{3}}(k^{(d)}_iD)=0,~~
J_{{\cal K}-\frac{1}{2}-\frac{n_1+n_2}{3}}(k^{(s)}_iD)=0.
\end{eqnarray}
we obtain the discrete set of wave numbers $k^{(u)}, k^{(d)}$, and $k^{(s)}$.
The orthogonal conditions are given by
\begin{eqnarray}
&&\int^D_0d\rho \rho J_\nu(k^{(p)}_i\rho)J_\nu(k^{(p)}_j\rho)=\int^D_0d\rho \rho J_{\nu\pm 1}(k^{(p)}_i\rho)J_{\nu\pm 1}(k^{(p)}_j\rho)
=\delta_{ij}\frac{D^2}{2}[J_{\nu\pm 1}(k^{(p)}_iD)]^2,~~
\end{eqnarray}
where
\begin{eqnarray}
\nu={\cal K}-\frac{1}{2}+\frac{2n_1-n_2}{3},{\cal K}-\frac{1}{2}-\frac{n_1-2n_2}{3},{\cal K}-\frac{1}{2}-\frac{n_1+n_2}{3}. \nonumber 
\end{eqnarray}
The eigenproblem can be solved numerically. If we take the entire infinite set of wave numbers (which
means an infinite size of the matrices), then the spectrum becomes exact. The normalization constants
for the basis vectors are
\begin{eqnarray}
&&{\cal N}^{(u)}_i=\biggl[\frac{2\pi D^2|\epsilon |}{|\epsilon|+m}\Bigl(J_{{\cal K}+\frac{1}{2}+\frac{2n_1-n_2}{3}}(k^{(u)}_iD)\Bigr)^2\biggr]^{-1/2}\,, \nonumber \\
&&{\cal N}^{(d)}_i=\biggl[\frac{2\pi D^2|\epsilon |}{|\epsilon|+m}\Bigl(J_{{\cal K}+\frac{1}{2}-\frac{n_1-2n_2}{3}}(k^{(d)}_iD)\Bigr)^2\biggr]^{-1/2}\,, \\
&&{\cal N}^{(s)}_i=\biggl[\frac{2\pi D^2|\epsilon |}{|\epsilon|+m}\Bigl(J_{{\cal K}+\frac{1}{2}-\frac{n_1+n_2}{3}}(k^{(s)}_iD)\Bigr)^2\biggr]^{-1/2}\,. \nonumber 
\end{eqnarray}

To obtain the spectral flow, we use the linearly interpolated field
\begin{eqnarray}
\Phi_{\rm intp}(x,\lambda)=(1-\lambda)\Phi_\infty+\lambda\Phi(x),~~~~0\leq \lambda \leq 1
\label{intpfield}
\end{eqnarray}
where $\Phi(x)$ is the field with a nontrivial topology and $\Phi_\infty(x)$ is the asymptotic field. 
In Fig. \ref{spectralflow}, we present
some typical plots of spectral flows. Some special energy levels pass from the negative to the positive
continuum. The number of fermionic levels crossing zero is always equal to the topological charge of the
field $\Phi(x)$. As is seen, these levels are normalizable modes, and their behavior is then described by the
index theorem.

The normalizable fermion density is given by (\ref{density}). Plots of it are shown in Fig. \ref{fermiondensity}. To save computer
time in the numerical calculation, we chose the cylinder radius $D = 20.0$ and $192$ points for discretizing the
momenta.

\section{Generalization to the higher $N$}

We have considered the cases $N = 1, 2$, but our analysis can be directly generalized to larger $N$. In
fact, soliton equations of motion (\ref{equationr}) are written for general $N$; moreover, some results can be found in~\cite{Amari:2015sva}.
We therefore concentrate on discussing how the fermionic part should be treated for larger $N$. It is easy to
show that for general $N$, the conserved quantum number commuting with Hamiltonian (\ref{hamiltonian}) is
\begin{eqnarray}
{\cal K}=\ell_3+\frac{\sigma_3}{2}+\sum_{k=1}^Nn_k\biggl[\frac{1}{3}I_{N+1\times N+1}-\tilde{I}_k\biggr]
\label{conservN}
\end{eqnarray}
where 
\begin{eqnarray}
\tilde{I}_k=
\left(\begin{array}{ccccc}
0 &  &  &  &  \\
  &\ddots &  &  &  \\  
  &&1&\leftarrow k&\\ 
  &&\uparrow &\ddots  &  \\ 
  &&k&  & 0 \\ 
\end{array}\right)
\end{eqnarray}
For $N=2$, of course (\ref{conservN}) coincides with (\ref{conserv2}). 

The plain wave basis can be written as
\begin{eqnarray}
&&\phi_{\cal K}^{(1)}(k_i^{(1)},\bm{x})={\cal N}^{(1)}_i
\left( 
	\begin{array}{c}
	\omega_{i,\epsilon}^{(1)-}J_{{\cal P}}(k_i^{(k)}\rho)e^{i{\cal P}\varphi}  
	\\
	\omega_{i,\epsilon}^{(1)+}J_{{\cal P}+1}(k_i^{(k)}\rho)e^{i({\cal P}+1)\varphi}
	\end{array}
\right)
	\otimes
\left.
\left(
	\begin{array}{c}
	1 \\\vdots \\0 \\\vdots \\0
	\end{array}
\right)~~\right\}
\,\text{\scriptsize $N$}\,,
	\\
&&\hspace{6cm}\vdots \nonumber \\
&&\phi_{\cal K}^{(k)}(k_i^{(k)},\bm{x})={\cal N}^{(k)}_i
\left( 
	\begin{array}{c}
	\omega_{i,\epsilon}^{(k)+}J_{{\cal P}-n_1+n_k}(k_i^{(k)}\rho)e^{i({\cal P}-n_1+n_k)\varphi}  
	\\
	\omega_{i,\epsilon}^{(k)-}J_{{\cal P}+1-n_1+n_k}(k_i^{(k)}\rho)e^{i({\cal P}+1-n_1+n_k)\varphi}
	\end{array}
\right)
	\otimes
\left(
	\begin{array}{c}
	0 \\\vdots \\1 \\\vdots \\0
	\end{array}
\right)
\,\leftarrow\text{\scriptsize $k$}\,,
	\\
&&\hspace{6cm}\vdots \hspace{4cm}k=2,\cdots,N-1\nonumber \\
&&\phi_{\cal K}^{(N)}(k_i^{(N)},\bm{x})={\cal N}^{(N)}_i
\left( 
	\begin{array}{c}
	\omega_{i,\epsilon}^{(N)+}J_{{\cal P}-n_1}(k_i^{(N)}\rho)e^{i({\cal P}-n_1)\varphi}  
	\\
	\omega_{i,\epsilon}^{(N)-}J_{{\cal P}+1-n_1}(k_i^{(N)}\rho)e^{i({\cal K}+1-n_1)\varphi}
	\end{array}
\right)
	\otimes
\left(
	\begin{array}{c}
	0 \\\vdots \\0 \\\vdots \\1
	\end{array}
\right)	\nonumber 
\end{eqnarray}
where 
\begin{eqnarray}
{\cal P}:={\cal K}-\frac{1}{2}+\frac{2n_1-\sum_{k=2}^Nn_k}{3}
\end{eqnarray}
and
\begin{eqnarray}
\omega_{i,\epsilon_>0}^{(p)+}=\omega_{i,\epsilon<0}^{(p)-}=1,~~~~
\omega_{i ,\epsilon>0}^{(p)-}=\omega_{i,\epsilon<0}^{(p)+}=\frac{-{\rm sgn}(\epsilon)k_i^{(p)}}{|\epsilon|+m},~~~~p=1,\cdots,N\,.
\end{eqnarray} 
The wave numbers $k^{(k)}$ are discretized by the boundary conditions
\begin{eqnarray}
J_{{\cal P}-n_1+n_k}(k^{(k)}_iD)=0,
J_{{\cal P}-n_1}(k^{(N)}_iD)=0,~~k=1,\cdots,N-1\,.
\end{eqnarray}
The orthogonal conditions are then
\begin{eqnarray}
&&\int^D_0d\rho \rho J_\nu(k^{(p)}_i\rho)J_\nu(k^{(p)}_j\rho)=\int^D_0d\rho \rho J_{\nu\pm 1}(k^{(p)}_i\rho)J_{\nu\pm 1}(k^{(p)}_j\rho)
=\delta_{ij}\frac{D^2}{2}[J_{\nu\pm 1}(k^{(p)}_iD)]^2\,,\\
&&\hspace{10cm}\nu={\cal P}-n_1+n_k,{\cal P}-n_1\,. \nonumber 
\end{eqnarray}
The normalization constants of the basis are then
\begin{eqnarray}
{\cal N}^{(k)}_i=\biggl[\frac{2\pi D^2|\epsilon |}{|\epsilon|+m}\Bigl(J_{{\cal P}+1-n_1+n_k}(k^{(k)}_iD)\Bigr)^2\biggr]^{-1/2}\,,~~
{\cal N}^{(N)}_i=\biggl[\frac{2\pi D^2|\epsilon |}{|\epsilon|+m}\Bigl(J_{{\cal P}+1-n_1}(k^{(N)}_iD)\Bigr)^2\biggr]^{-1/2}\,. 
\end{eqnarray}
In terms of the basis, 
the analysis of Eq.(\ref{secular_equation}) can be solved numerically.

\section{Summary}

We have studied the spectrum of the fermions coupled with the baby-Skyrme model or the $\mathbb{C}P^N$ Skyrme-Faddeev-type model. 
We computed the spectral flow of the fermionic one-particle state giving
the level-crossing picture. The baby skyrmions are assumed to be anyons because $\Pi_4(\mathbb{C}P^1)$ is trivial. But
the anyon angle $\Theta=n_c\pi{\rm sgn}(m)$ should be an integer corresponding to the number $Q_{\rm top}$ of normalizable
modes of fermions. Solutions of the $\mathbb{C}P^N$ model are anyons because the Hopf term is perturbative. On
the other hand, $Q_{\rm top}$ normalizable states are localized on the soliton. Hence, in the case of the target
space $\mathbb{C}P^N$, there is an inconsistency between the statistical natures of fermions and solitons. Perhaps, this
inconsistency can be overcome; we will devote our next paper to this.

\vspace{1cm}

\noindent {\bf Acknowledgments} 
N.S. would like to thank the conference organizers of MQFT 2018 for kind
accommodation and hospitality. 

\vspace{1cm}

\noindent {\bf Appendix A}

In this appendix, we use the notation in \cite{Diakonov:1987ty}. The partition function is given by the integral
\begin{eqnarray}
\mathcal{Z}=\int \mathcal{D}\psi\mathcal{D}\bar{\psi} e^{\int d^3x \bar{\psi}i\mathcal{D}_A\psi}
=(\det i\mathcal{D}_A)^{n_c}
\end{eqnarray}
where $\psi$ and $\bar\psi$ are Dirac fields and $i\mathcal{D}_A=i\gamma_\alpha(\partial_\alpha-iA_\alpha)-mU$ where $A_\alpha$ is a U(1) gauge field. 
The gamma matrices are defined by $\gamma_\alpha=-i\sigma_\alpha$. 
The effective action $\omega=n_c\ln \det i\mathcal{D}_A$ is split in its real and imaginary part
\begin{eqnarray}
\omega_{\rm Re}=\frac{n_c}{2}\ln \det \mathcal{D}_A^\dagger \mathcal{D}_A, ~~~~
\omega_{\rm Im}=\frac{n_c}{2i}\ln \det\frac{i\mathcal{D}_A}{-i\mathcal{D}_A^\dagger}\,.
\end{eqnarray}
It is easy to see that as $A_\mu\to 0$ (and $\mathcal{D}_A\to \mathcal{D}$), 
\begin{eqnarray}
\mathcal{D}^\dagger \mathcal{D}=-\partial^2+m^2+im\gamma_\alpha\partial_\alpha U,~~~~
\mathcal{D}\mathcal{D}^\dagger =-\partial^2+m^2-im\gamma_\alpha\partial_\alpha U\,.
\end{eqnarray}
Appropriately subtracting the vacuum state $U = 1$, we obtain
\begin{eqnarray}
\omega_{\rm Re}&=&\frac{n_c}{2}{\rm Sp}{\rm ln}\Bigl(1+\frac{im\gamma_\alpha \partial_\alpha U}{-\partial^2+m^2}\Bigr)
=\frac{n_c}{2}\int d^3x\int \frac{d^3k}{(2\pi)^3}e^{-ikx}{\rm Trln}\Bigl(1+\frac{im\gamma_\alpha \partial_\alpha U}{-\partial^2+m^2}\Bigr)e^{ikx} \nonumber \\
&&\hspace{3cm}=\frac{n_c}{2}\int d^3x\int \frac{d^3k}{(2\pi)^3}{\rm Trln}\Bigl(1+\frac{im\gamma_\alpha \partial_\alpha U}{k^2+m^2-2ik\partial-\partial^2}\Bigr)
\end{eqnarray}
Expanding the above expression in powers of $2ik\partial+\partial^2$ and $\gamma_\alpha \partial_\alpha U$ one gets in the lowest non-zero term
\begin{eqnarray}
 \omega^{(2)}_{\rm Re}=-n_c\frac{|m|}{16\pi}\int d^3x {\rm Tr}(\partial_\mu U\partial_\mu U)\,.
\end{eqnarray}
After taking the spinor trace and switching to the Minkowski metric one gets
the action (\ref{effectiveaction_fr}). 

Taking the variation $\mathcal{D}\to\mathcal{D}+\delta\mathcal{D}, \mathcal{D}^\dagger\to \mathcal{D}^\dagger+\delta\mathcal{D}^\dagger$ 
for $A_\mu\to 0$ the imaginary part is
\begin{eqnarray}
\delta\omega_{\rm Im}=\frac{n_c}{2i}
{\rm Tr}\biggl(\frac{1}{\mathcal{D}^\dagger \mathcal{D}}\mathcal{D}^\dagger \delta \mathcal{D}
-\frac{1}{\mathcal{D}\mathcal{D}^\dagger}\mathcal{D} \delta \mathcal{D}^\dagger\biggr).
\end{eqnarray}
It contains the product of three derivatives:
\begin{eqnarray}
\delta \omega_{\rm Im}^{(3)}=-n_c\frac{{\rm sgn}(m)}{32\pi}\int d^3x \epsilon^{\mu\nu\lambda}
{\rm Tr}(\partial_\mu U\partial_\nu U\partial_\lambda U U\delta U)\,.
\end{eqnarray}
In terms of new variable $a_{\mu}:=-iZ^\dagger \partial_\mu Z$, we can write the last formula as
\begin{eqnarray}
\delta \omega^{(3)}_{\rm Im}=n_c\frac{{\rm sgn}(m)}{2\pi}\int d^3x \epsilon^{\mu\nu\lambda}\delta a_\mu \partial_\nu a_\lambda\,.
\end{eqnarray}
It can be shown that this expression coincides with the variation of action (\ref{hopf}). For $A_\mu = 0$, 
we also have the two-derivative component
\begin{eqnarray}
\delta \omega^{(2)}_{\rm Im}|_{A_\mu\neq 0}
=-\frac{n_c}{16\pi i}\delta A_\mu \int d^3x\epsilon^{\mu\nu\lambda}
{\rm Tr}(\partial_\nu U\partial_\lambda U U)\,.
\end{eqnarray}
These two components contribute to the final expression  (\ref{effectiveaction_fi}).

\section*{References}
\bibliography{cpnfermion}

\end{document}